\documentclass[trackchanges,twocolumn, twocolappendix, appendixfloats]{aastex631} 
\usepackage{amsmath}
\usepackage{hyperref}
\usepackage{csvsimple}
\usepackage{verbatim}
\usepackage{graphicx}
\usepackage{booktabs}

\usepackage[english]{babel}
\usepackage[autostyle, english=american]{csquotes}
\usepackage{multirow}
\usepackage{colortbl}
\usepackage{xcolor}
\usepackage{calc}
\usepackage{xspace}
\shorttitle{Signatures of Feedback in LMC Star-Forming Regions}
\shortauthors{A. Green et al.}
\graphicspath{{./}{figures/}}
\newcommand{\avgm}{$\langle 8 \micron \rangle$\xspace}
\newcommand{\avgH}{$\langle \text{H} \alpha \rangle$\xspace}
\newcommand{\avgu}{$\langle u_{\nu} \rangle$\xspace}
\newcommand{\avgS}{$\langle \Sigma \rangle$\xspace}

\begin{document}

\defcitealias{solomon_mass_1987}{S87}

\title{The Size-Linewidth Relation and Signatures of Feedback from Quiescent to Active Star Forming Regions in the LMC}

\author[0000-0002-8432-3362]{Alex Green}
\affiliation{Department of Astronomy, University of Illinois, Urbana, IL 61801, USA}

\author{Tony Wong}
\affiliation{Department of Astronomy, University of Illinois, Urbana, IL 61801, USA}

\author{Rémy Indebetouw}
\affiliation{Department of Astronomy, University of Virginia, Charlottesville, VA, USA}
\affiliation{National Radio Astronomy Observatory, Charlottesville, VA, USA}

\author{Omnarayani Nayak}
\affil{Space Telescope Science Institute, 3700 San Martin Drive, Baltimore, MD 21218, USA}

\author{Alberto Bolatto}
\affiliation{Department of Astronomy, University of Maryland, College Park, MD, USA}

\author{Elizabeth Tarantino}
\affiliation{Department of Astronomy, University of Maryland, College Park, MD, USA}
\affil{Space Telescope Science Institute, 3700 San Martin Drive, Baltimore, MD 21218, USA}

\author[0000-0002-5307-5941]{Mónica Rubio}
\affiliation{Departamento de Astronomía,  Universidad de Chile, Santiago, Chile}

\author{Suzanne C. Madden}
\affiliation{Département d’Astrophysique, Université Paris-Saclay, Gif-sur-Yvette, France}

\author[0000-0002-2954-8622]{Alec S.\ Hirschauer}
\affil{Space Telescope Science Institute, 3700 San Martin Drive, Baltimore, MD 21218, USA}

\begin{abstract}
To investigate the effects of stellar feedback on the gravitational state of giant molecular clouds (GMCs), we study $^{12}$CO and $^{13}$CO ALMA maps of nine GMCs distributed throughout the Large Magellanic Cloud (LMC), the nearest star-forming galaxy to our own. We perform noise and resolution matching on the sample, working at a common resolution of 3.5 arcseconds (0.85 pc at the LMC distance of 50 kpc), and use the \textit{SCIMES} clustering algorithm to identify discrete substructure, or ``clumps." We supplement these data with three tracers of recent star formation: $8\micron$ surface brightness, continuum-subtracted H$\alpha$ flux, and interstellar radiation field energy density inferred from dust emission. The $^{12}$CO clumps identified cover a range of 3.6 dex in luminosity-based mass and 2.4 dex in average $8\micron$ surface brightness, representative of the wide range of conditions of the interstellar medium in the LMC. Our observations suggest evidence for increased turbulence in these clouds. While the turbulent linewidths are correlated with clump surface density, in agreement with previous observations, we find even better correlation with the three star formation activity tracers considered, suggesting stellar energy injection plays a significant role in the dynamical state of the clumps. The excess linewidths we measure do not appear to result from opacity broadening. $^{12}$CO clumps are found to be typically less gravitationally bound than $^{13}$CO clumps, with some evidence of the kinetic-to-gravitational potential energy ratio increasing with star-formation tracers. Further multi-line analysis may better constrain the assumptions made in these calculations. 
\end{abstract}
\section{Introduction}
Giant molecular clouds (GMCs), which range from tens to hundreds of parsecs and between $10^4$ and $10^7$ M$_{\odot}$ \citep{fukui_molecular_2010, heyer_molecular_2015}, are the sites of star formation. Star formation is inefficient in changing molecular gas mass into stellar mass. The star formation efficiency per free fall time has been measured in the Milky Way and other galaxies to be typically of order 1\% \citep{lee_observational_2016, ochsendorf_what_2017, utomo_star_2018}, indicating the inefficiency of star formation in typical interstellar environments, compared to an idealized state of unimpeded collapse. After stars form, they can alter their natal environments significantly enough to affect star formation rates (SFR) in the cloud. The process by which star formation enhances or suppresses local star formation is called stellar ``feedback," and includes heating and ionization from radiation \citep{gritschneder_driving_2009}, stellar winds and outflows \citep{haid_relative_2018, bally_protostellar_2016}, and supernovae \citep{walch_silcc_2015, koo_radiative_2020}. Results from simulations of galaxies demonstrate that stellar feedback is necessary for reproducing observed star formation efficiencies \citep{hopkins_self-regulated_2011} and is a key component in setting the interstellar medium (ISM) conditions, like increasing the temperature and decreasing the density of gas, that allow supernovae to drive outflows (\citealp{wise_birth_2012, emerick_stellar_2018}). Simulations that resolve GMCs show feedback processes driving small-scale turbulence \citep{gritschneder_driving_2009} and eventually dispersing the gas \citep{grudic_dynamics_2022}. Observational studies have found connections between such feedback processes---including stellar winds (\citealp{pabst_disruption_2019, tiwari_sofia_2021}), ionizing luminosity \citep{murray_star_2011}, and protostellar outflows \citep{kavak_dents_2022}---and the surrounding molecular and atomic gas (see also \citealp{barnes_comparing_2021} for a comparison of feedback sources).

Feedback has also been invoked as a possible source of energy to maintain the turbulence in GMCs. GMCs, both galactic and extragalactic, have been found to exhibit a size-linewidth relation ($R-\sigma_v$ relation), $\sigma_v \propto R^{1/2}$ (\citealp{larson_turbulence_1981, solomon_mass_1987}, hereafter \citetalias{solomon_mass_1987}; \citealp{bolatto_resolved_2008}). This relationship holds over several orders of magnitude in size, and is consistent for a large variety of clouds found in different environments. However, there is significant scatter in the $R-\sigma_v$ relation and debate as to whether a single power-law index holds for all size scales \citep{hennebelle_turbulent_2012}. 

 Measurement of the size-linewidth relation can suffer from observational biases. When inferred from low-density tracers, the slope can be measured to be flatter due to line-of-sight confusion of small clumps causing artificially higher observed velocity dispersions \citep{ballesteros_paredes_physical_2002}. The size-linewidth relation can be quite scattered when observed with $^{13}$CO \citep{simon_structure_2001}, and some studies using $^{12}$CO find no linear correlation \citep{schneider_bias_2004}. The size of structures studied in both \cite{simon_structure_2001} and \cite{schneider_bias_2004} span one order of magnitude, while the \citetalias{solomon_mass_1987} relation was established using observations spanning two orders of magnitude in size. As noted by \cite{ballesteros_paredes_physical_2002}, a larger dynamic range in observations is necessary to accurately recover the slope. 

The $R-\sigma_v$ relation has been explained as the manifestation of supersonic, or Burgers, turbulence \citep{mckee_theory_2007}. Turbulence at the cloud scale is driven by large-scale flows arising from the galactic gravitational potential \citep{klessen_accretion-driven_2010}. However, as suggested by \cite{mestel_star_1956}, due to loss of energy from gas compression and radiation, turbulence cannot support a cloud from collapse for longer than a few dynamical times. Therefore, if the $R-\sigma_v$ relation is due to turbulence, a mechanism for energy injection on the scales relevant to the relation is required. The feedback sources discussed above could be responsible mechanisms.

Through a $^{13}$CO survey of molecular clouds along the Galactic Plane, \cite{heyer_re-examining_2009} studied the clouds used in \citetalias{solomon_mass_1987} at higher resolution and found that the normalization of the fiducial size-linewidth relation
\begin{equation} \label{eq:nu0_def}
    \nu_0 \equiv \frac{\sigma_v}{R^{1/2}}
\end{equation}
was related to the cloud surface density as $\nu_0 \propto \Sigma^{1/2}$. This scaling is consistent with the condition for virial equilibrium of spherical, uniform density clouds, 
\begin{equation}
    \nu_0 = \left(\pi G \Sigma /5 \right)^{1/2},
\end{equation}
and Larson's scaling relations predict clouds to be in virial equilibrium. However, the data presented in \cite{heyer_re-examining_2009} are offset from the virial equilibrium condition, suggesting clouds have kinetic energies in excess of simple virial equilbrium. Other models in which the observed linewidths are gravity-driven have been proposed and can explain some of the offset from virial equilibrium. These include the hierarchical collapse, with clouds predicted to be in energy equipartition \citep{ballesteros-paredes_gravity_2011}, as well as clouds confined by external pressure \citep{field_does_2011}. The origins of these intercloud pressures are unclear, but there could be contributions exerted by stellar feedback sources. Despite considerations from such models, there remain data with $\nu_0$ in excess of model predictions. 

As the closest star-forming galaxy to the Milky Way (50 kpc;  \citealp{pietrzynski_distance_2019}), the Large Magellanic Cloud (LMC) allows for the most detailed studies of extragalactic molecular clouds. The LMC has a total mass of $6\times10^9 M_{\odot}$ \citep{1990A&ARv...2...29W}, and with metallicity half of the solar neighborhood value, the LMC may also serve as a proxy for star formation at higher redshift (\citealp{westerlund_magellanic_1997, 2010ARA&A..48..547F}). Recent studies in the LMC have observed a $R-\sigma_v$ relation to hold for substructure within GMCs, but have found that the normalization of  the relation (i.e., the linewidth at a given size) varies with local and cloud-scale measures of the radiation field \citep{wong_alma_2017, wong_relations_2019}. These results suggest that the $R-\sigma_v$ relation is related to turbulence driven in part by local stellar feedback. 

In order to study the contribution of stellar feedback to the $R-\sigma_v$ relation in GMCs, we assemble a matched, high-resolution sample of nine clouds in the LMC, covering a large range of CO brightness, extending the sample studied in \cite{wong_relations_2019} and using newly-published observations of 30 Doradus covering a larger field of view than previously mapped \citep{wong_30_2022}. We employ 8$\micron$, H$\alpha$, and interstellar radiation field energy density to understand drivers of excess linewidths in the molecular gas. This paper is organized as follows. In Section \ref{sec:methods}, we describe the archival ALMA observations of CO and ancillary observations of star-formation tracers. In Section \ref{sec:results}, we describe our results fitting star-formation tracers to the residuals of the $R-\sigma_v$ relations of the sample. We discuss these results in Section \ref{sec:disc}. We make concluding remarks in Section \ref{sec:concl}.
\begin{figure}[t] 
    \centering
    \includegraphics[trim={0.82in 0.175in 0.325in 0.655in}, clip, width=0.49\textwidth]{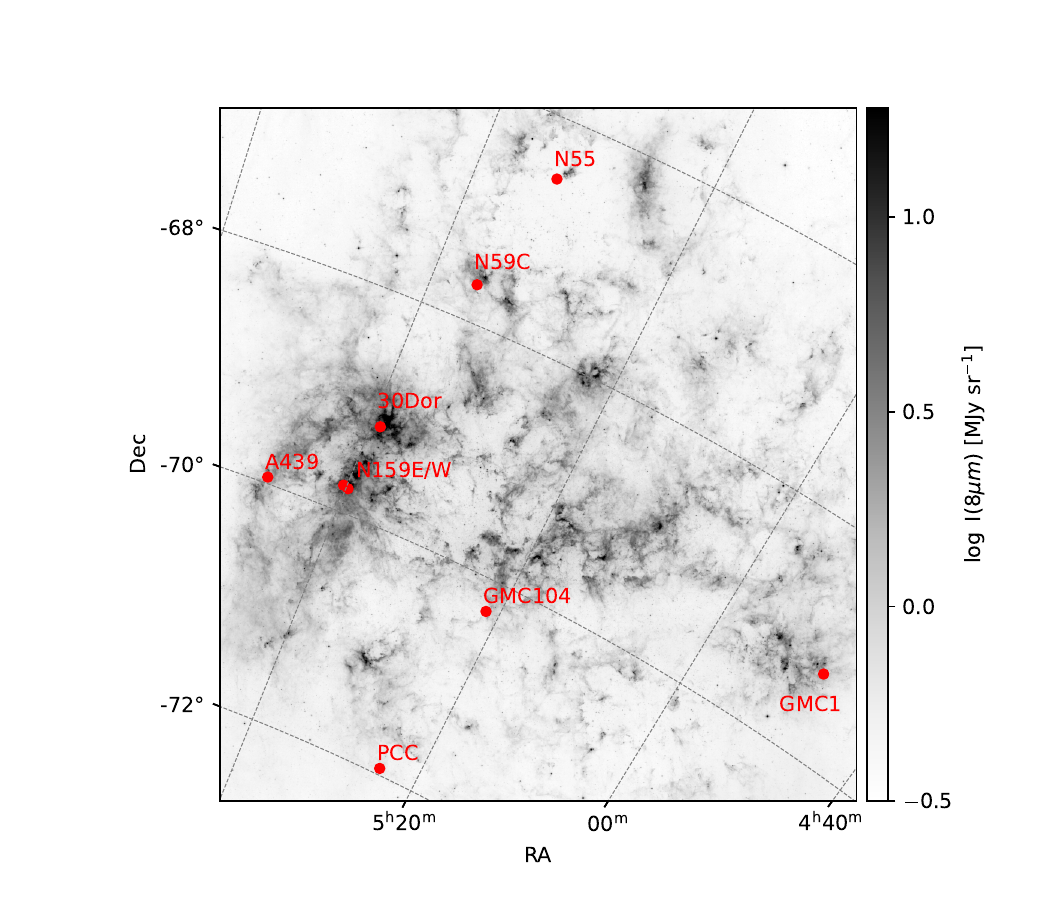}
    \caption{LMC SAGE $8\micron$ map from \cite{meixner_spitzer_2006} with locations of the nine clouds in the sample (red points). The sample consists of regions spanning a considerable range of CO- and MIR-brightness across the LMC. We compute a noise level in eight signal-free regions in order to estimate uncertainties in measurements of 8$\mu$m. These regions contain $\sim 5.3\times 10^{7}$ pixels at the original IRAC channel 4 resolution, around 9.6\% of the map.}
    \label{fig:LMC_overview}
\end{figure}

\begin{figure}[ht] 
    \centering
    \includegraphics[trim={0 0.2in 0.385in 0.615in}, clip, width=\columnwidth]{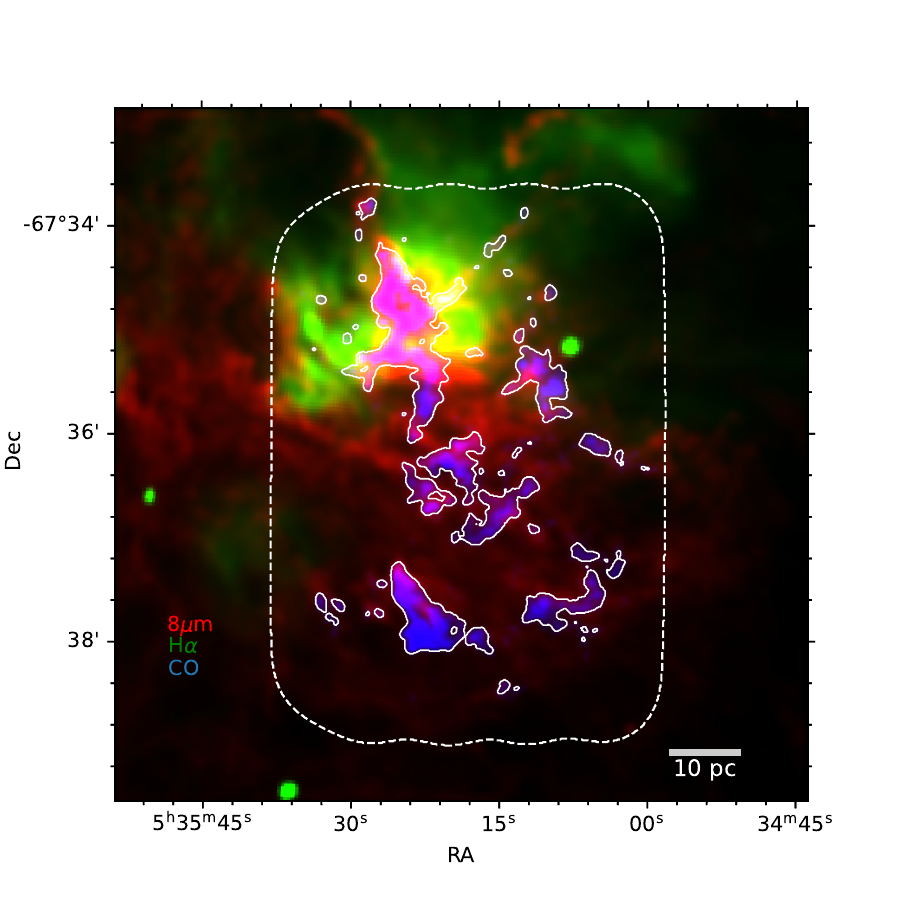}
    \caption{Three-color image of N59C in $^{12}$CO moment 0 (blue, 5 K km s$^{-1}$ in solid white contours), IRAC 8$\micron$ (red, convolved to 3\farcs5), and MCELS continuum-subtracted H$\alpha$ (green). The dashed white line marks the ALMA field of view. The northern CO-bright region of the cloud is also H$\alpha$- and 8$\micron$-bright, but H$\alpha$, which is not corrected for extinction here, does not trace the southern $^{12}$CO emission. While the CO and 8$\micron$ do not follow one-to-one spatial correspondence, they do match at coarser scales over the extent of the cloud. The 8$\micron$ emission is also bright in regions where CO is not detected at high significance with these observations. The blue, red, and green channels are clipped at the 93$^{\text{rd}}$, 98$^{\text{th}}$, and 99$^{\text{th}}$ percentile of the data.}
    \label{fig:N59C_RGB}
\end{figure}

\section{Methods} \label{sec:methods}
\subsection{Resolution-Matching Archival \texorpdfstring{$^{12}$CO}{12CO} and \texorpdfstring{$^{13}$CO}{13CO} Observations} \label{sec:resmatching}
We base this study on published Atacama Large Millimeter/submillimeter Array (ALMA) $^{12}$CO and $^{13}$CO observations. Each molecular cloud analyzed in our sample has been observed in both $^{12}$CO and $^{13}$CO in either the J = 1 -- 0 (ALMA Band 3) or J = 2 -- 1 (ALMA Band 6) transition. 
Observations of A439, GMC1, GMC104, N59C in (1 -- 0) and PCC in (2 -- 1) were obtained from \cite{wong_relations_2019}, N159E and N159W in (2 -- 1) from \cite{fukui_high-mass_2015} and \cite{saigo_kinematic_2017}, N55 in (1 -- 0) from \cite{2018ApJ...853..175N}, and 30 Doradus in (2 -- 1) from \cite{wong_30_2022}. We match all the observations to a common channel resolution of 200 m s$^{-1}$ and spatial resolution of 3\farcs 5. At the adopted LMC distance of 50 kpc, this corresponds to 0.85 pc. We match all of the spectral cubes of a given line to a common noise level. We first estimate the noise by taking the standard deviation from the median absolute deviation (\texttt{astropy.stats.mad\_std}) in 20 edge channels of each cloud ($\sigma_{\text{cloud}}$). We then choose the common noise by finding the maximum value of the nine clouds for each line ($\sigma_{\text{target}}$). We convolve a cube of Gaussian noise ($\sigma_{\text{GN}}$)  with the beam and normalize it to have an RMS of $\sqrt{\sigma_{\text{target}}^2 - \sigma_{\text{GN}}^2}$. This scaled cube is then added to the image cubes. The peak brightness and noise levels of the cubes are presented in Table \ref{table:overview_table} and the locations of the clouds are presented in Figure \ref{fig:LMC_overview} with the $8\micron$ map of the LMC.

The spectral response of the instrument affects the inferred width of emission with line profile comparable to the channel width. We remove this response by performing deconvolution-in-quadrature of the linewidth, following \cite{rosolowsky_bias-free_2006}:
\begin{equation}
    \sigma_{v, \text{deconvolved}} = \sqrt{\sigma_{v, \text{measured}}^2 - \sigma_{v, \text{response}}^2}
\end{equation}
We use the spectral response prescribed by \cite{leroy_portrait_2016} to account for correlations in adjacent channels, 
\begin{equation}
    \sigma_{v, \text{response}} = \left(\Delta \nu_{\text{channel}}/\sqrt{2\pi}\right) \left(1.0 + 1.18 k + 10.4 k^2\right)
\end{equation}
where $k$ is related to the correlation coefficient between noise in successive channels in the cube with signal masked, 
\begin{equation}
    k = 0.47r - 0.23r^2 - 0.16r^3 + 0.43r^4\;.
\end{equation}
All of the clouds yield $r < 0.65$, within the range in which $r$ can be used to determine the parameter of the window function specified by \cite{leroy_portrait_2016}. The channel-to-channel correlation coefficient is calculated using the \texttt{SpectralCubeTools}\footnote{\url{https://github.com/astrojysun/SpectralCubeTools}} package (\citealp{sun_cloud-scale_2018, sun_molecular_2020}). We mask the signal using dilated masks produced using the \texttt{maskmoment}\footnote{\url{https://github.com/tonywong94/maskmoment}} package. We use the default masking parameters except \texttt{snr\_hi\_minch = 2} and \texttt{minbeam = 2}, requiring that the 4$\sigma$ contour mask span 2 channels at all pixels, and that all masked regions have a minimum area of 2 beams. The 5 K km s$^{-1}$ contour level of the $^{12}$CO moment 0 is shown in Figure \ref{fig:N59C_RGB}, and the $^{12}$CO moment 0 maps of all the clouds in the sample are shown in Figure \ref{fig:scimes_overview}.

\begin{deluxetable*}{lcrrrccrrrrrr}[t] \label{table:overview_table}
\rotate
\tablehead{\colhead{Region} & \colhead{Transition} & \colhead{\hspace{0.5cm}R.A.} & \colhead{\hspace{0.5cm}Dec.} & \colhead{Field size} & \colhead{$^{12}$CO beam size} & \colhead{Velocity resolution} & \colhead{T$_{\text{rms}, 12}$} & \colhead{T$_{\text{peak}, 12}$} & \colhead{T$_{\text{rms}, 13}$} & \colhead{T$_{\text{peak}, 13}$} & \colhead{Reference}\\ \colhead{ } & \colhead{J = } & \colhead{\hspace{0.5cm}$\mathrm{h:m:s}$} & \colhead{\hspace{0.5cm}$\mathrm{d:m:s}$} & \colhead{arcmin$^2$} & \colhead{arcsec $\times$ arcsec} & \colhead{km s$^{-1}$} & \colhead{$\mathrm{K}$} & \colhead{$\mathrm{K}$} & \colhead{$\mathrm{K}$} & \colhead{$\mathrm{K}$} & \colhead{ }}
 \tablecaption{Summary of the clouds in the sample. The coordinates are at the center of each mosaic. The beam size and velocity resolution are the values at which each cloud was originally imaged (see citations in the Reference column), while the rms and peak temperatures are computed after convolution to the common 3\farcs 5 beam and 200 m s$^{-1}$ channel width.}
\startdata
30Dor & (2-1) & \hspace{0.5cm}5:38:36 & \hspace{0.5cm}-69:05:27 & 34.147 & 1.75 $\times$ 1.75 & 0.1 & 0.779 & 34.868 & 0.251 & 16.245 & (1) \\
A439 & (1-0) & \hspace{0.5cm}5:47:26 & \hspace{0.5cm}-69:52:46 & 19.172 & 2.10 $\times$ 1.53 & 0.2 & 1.049 & 14.176 & 0.250 & 4.789 & (2) \\
GMC1 & (1-0) & \hspace{0.5cm}4:47:40 & \hspace{0.5cm}-69:09:42 & 36.976 & 1.89 $\times$ 1.64 & 0.2 & 0.858 & 14.251 & 0.303 & 5.265 & (2) \\
GMC104 & (1-0) & \hspace{0.5cm}5:21:04 & \hspace{0.5cm}-70:13:29 & 26.186 & 2.46 $\times$ 2.28 & 0.2 & 1.240 & 17.219 & 0.337 & 6.499 & (2) \\
N59C & (1-0) & \hspace{0.5cm}5:35:19 & \hspace{0.5cm}-67:36:12 & 35.670 & 2.12 $\times$ 1.74 & 0.2 & 0.818 & 27.243 & 0.304 & 6.247 & (2) \\
N159E & (2-1) & \hspace{0.5cm}5:40:09 & \hspace{0.5cm}-69:44:45 & 17.093 & 1.21 $\times$ 0.84 & 0.2 & 0.664 & 7.303 & 0.264 & 2.855 & (3) \\
N159W & (2-1) & \hspace{0.5cm}5:39:37 & \hspace{0.5cm}-69:45:49 & 14.952 &  1.35 $\times$ 0.86 & 0.2 & 0.742 & 8.028 & 0.286 & 3.889 & (3) \\
N55 & (1-0) & \hspace{0.5cm}5:32:15 & \hspace{0.5cm}-66:26:14 & 68.467 & 3.5 $\times$ 2.3 & 0.2 & 0.699 & 24.940 & 0.292 & 4.421 & (4) \\
PCC & (2-1) & \hspace{0.5cm}5:23:56 & \hspace{0.5cm}-71:54:40 & 16.152 &  1.81 $\times$ 1.24 & 0.2 & 0.699 & 11.135 & 0.245 & 3.843 & (5)
\enddata
\tablecomments{(1): \cite{wong_30_2022}; (2): \cite{wong_relations_2019}; (3): \cite{nayak_molecular_2018}; (4): \cite{naslim_alma_2018}; (5): \cite{wong_alma_2017}}
\end{deluxetable*}

\begin{deluxetable}{lrrrr}[t] \label{table:cloud_size}
\tablehead{\colhead{Region} & \multicolumn{2}{c}{$^{12}$CO} & \multicolumn{2}{c}{$^{13}$CO} \\
\colhead{} & \colhead{Mass} & \colhead{Area} & \colhead{Mass} & \colhead{Area} \\
\colhead{} & \colhead{[$10^4$ M$_{\odot}$]} & \colhead{[arcmin$^2$]} & \colhead{[$10^4$ M$_{\odot}$]} & \colhead{[arcmin$^2$]}}
 \tablecaption{Total mass and area covered by $^{12}$CO and $^{13}$CO \textit{SCIMES} clumps in each cloud. }
\startdata
30Dor & 26.298 & 6.098 & 12.974 & 2.478 \\
A439 & 4.941 & 2.179 & 1.196 & 0.639 \\
GMC1 & 2.750 & 2.006 & 1.265 & 0.899 \\
GMC104 & 4.375 & 1.937 & 1.332 & 0.784 \\
N59C & 7.724 & 3.843 & 1.931 & 0.525 \\
N159E & 0.497 & 0.582 & 0.209 & 0.325 \\
N159W & 0.921 & 0.631 & 0.376 & 0.554 \\
N55 & 3.387 & 1.345 & 0.082 & 0.045 \\
PCC & 1.607 & 1.899 & 0.330 & 0.471
\enddata
\end{deluxetable}

\subsection{Segmenting CO Emission} \label{sec:COsegmentation}
\texttt{astrodendro}\footnote{\url{https://dendrograms.readthedocs.io}} is a Python package which segments position-position-velocity (PPV) data into hierarchical structures called ``leaves," ``branches," and ``trunks." These structures are nested, so leaves are fully contained in branches, and branches in trunks. For each cloud, we estimate the noise using the median absolute deviation ($\sigma_{\text{MAD}}$) of the pixels not included in the moment mask. Independent structures are only identified for pixels with brightness temperature above $3\sigma_{\text{MAD}}$ (\texttt{min\_value}) and with a size of at least two beams (\texttt{min\_npix}). We require that two local emission peaks must have a difference of at least 2.5$\sigma_{\text{MAD}}$ to be considered independent leaves (\texttt{min\_delta}). Position and flux of structures are calculated, and linewidth and size along structure major and minor axes ($\sigma_v, \sigma_{\text{maj}}, \sigma_{\text{min}}$) are measured from the second moment along each axis. Following \citetalias{solomon_mass_1987}, the structure radius is calculated as $R = 1.91\sqrt{\sigma_{\text{maj}} \sigma_{\text{min}}}$. \cite{2008ApJ...679.1338R} introduce a ``bijection" approach for dendrogram decomposition which we follow: All of the emission bound by a dendrogram structure (e.g., structure A) is considered to be from that structure, rather than subtracting the emission contributed by a parent structure (e.g., structure B such that $A \in B$) as in the ``clipping" approach also described. Once the dendrogram has been computed for a PPV data cube, we create a catalog of dendrogram structure properties and errors. 

\cite{rosolowsky_bias-free_2006} use a bootstrapping method to compute error. In this approach, the dendrogram structure masks are preserved, and the properties of the mask are recalculated after a random shuffling of pixels in the mask. This is repeated for many realizations in each structure, and the property error is calculated as the standard deviation of the realizations. We use bootstrapping with 100 iterations. In our sample, there are between 3.5 and 25 pixels per beam width. To account for correlated pixels, \cite{rosolowsky_bias-free_2006} scale the uncertainty by the square root of this oversampling rate. We assume that noise in two adjacent channels is correlated, so we instead scale the uncertainty by the square root of twice the number of pixels in the beam. This is a conservative estimate, and in some of the clouds studied, this leads to overestimated uncertainties since the native correlator resolutions of the observations used range from 40 m s$^{-1}$ to 184 m s$^{-1}$. To study the robustness of this method to number of realizations, we compute property errors using a number of bootstrapping iterations ranging from 50 to 400. We find that errors are robust after $\sim100$ realizations of a structure. We perform another method of error estimations by adding noise to the PPV cube. This method is similar in using the standard deviation of many new realizations of the data, but for each realization, convolved noise with RMS equal to the existing noise is added to the entire cube rather than shuffling the pixels in individual structures. In either case, we assume the new realization of the data does not change the dendrogram segmentation of the cube (i.e., we apply fixed dendrogram masks when computing properties). For $\sim$100 realizations, the noise-addition method returns error measures consistent with bootstrapping, so we proceeded to use the bootstrapping method to derive dendrogram property uncertainties. We approximate the iso-emission surfaces as 2D gaussians, and then recalculate properties in the structure by taking the ellipse formed by the clump major and minor axis and deconvolving by the beam. We impose a minimum fractional uncertainty of 5\% on the measured properties of the dendrogram structures.  

With this catalog of dendrogram structures, we use the Spectral Clustering for Molecular Emission Segmentation (\textit{SCIMES})\footnote{\url{https://scimes.readthedocs.io}} algorithm to gather structures into ``clumps," associations based on PPV volume \citep{colombo_graph-based_2015}. Since many of the leaves of the dendrogram are comparable to the beam size, this set of clumps provides us with a coarser decomposition of the data that can allow us to investigate environmental effects on the sub-cloud and cloud scale while retaining a fairly high number of independent structures (117 clumps among all of the GMCs in our sample in CO, 50 in $^{13}$CO). While many of the leaves in the dendrogram are enclosed within common parent structures which are identified as individual clusters by \textit{SCIMES}, sometimes leaves are identified as independent structures. \textit{SCIMES} can be made to consider these independent leaves as clusters by using the \texttt{save\_isol\_leaves} keyword in the \texttt{SpectralCloudstering} class. The isolated leaves constitute a population of small radius, but still significantly detected emission, so we include them in our sample. 

\begin{figure*}[ht] 
    \centering
    \includegraphics[width=\textwidth,trim={0 0 0 0},clip]{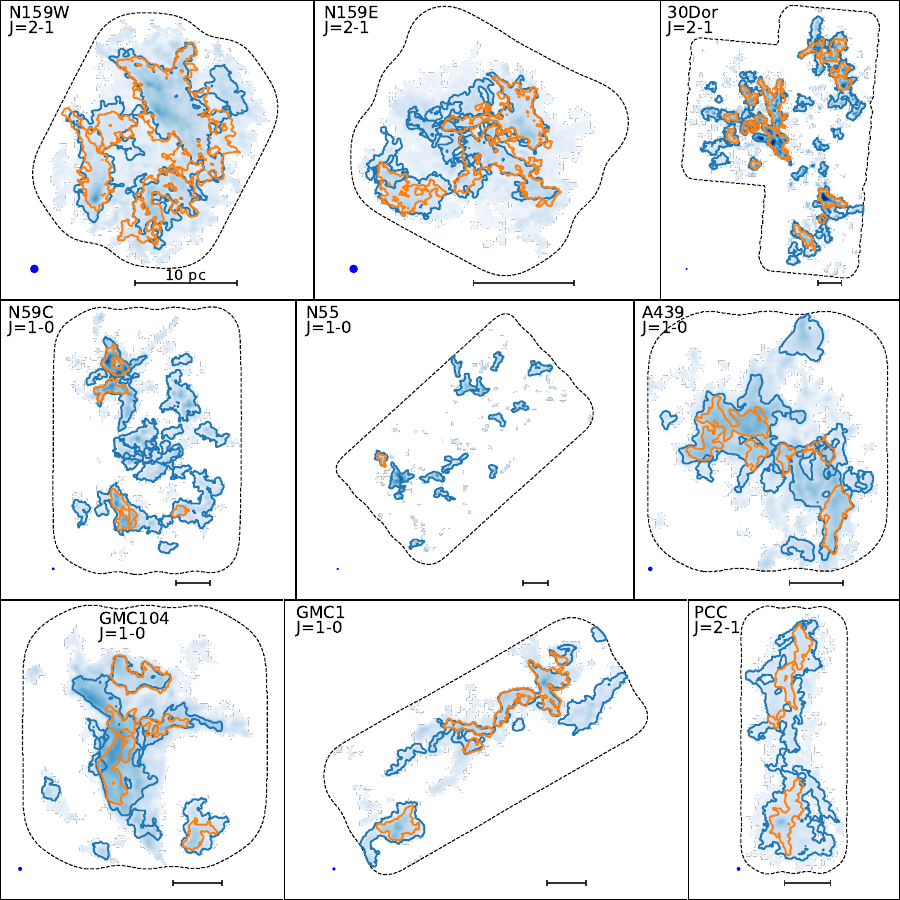}
    \caption{Projected \textit{SCIMES} $^{12}$CO (blue contours) and $^{13}$CO (orange contours) clumps on the $^{12}$CO moment 0 maps with square root stretch from 0 to 175 K km s$^{-1}$. Dashed lines mark the ALMA field of view of CO, with the beam and 10 pc scale at the bottom of each map. While the clumps cover a large range of scales of structure in the clouds, they do not contain all of the CO emission. Some clumps overlap in cases where there is emission with different velocities along the line of sight.}
    \label{fig:scimes_overview}
\end{figure*}

Because deconvolution can introduce errors on size, we introduce a method to remove high-uncertainty clumps from the sample. \texttt{astrodendro} provides an ellipsoid approximation to the dendrogram structure boundaries. We impose a cut that removes clumps with minor axis less than $\sqrt{2}$ times the beam width.  Our final selection of clumps, plotted as contours over the $^{12}$CO moment 0 in Figure \ref{fig:scimes_overview}, covers an order of magnitude in deconvolved linewidths, and more than an order of magnitude in deconvolved radii (Figure \ref{fig:histograms}). The effect of linewidth deconvolution on clumps is small, with the largest decrease in $^{12}$CO linewidth being by 4\% (9\% for $^{13}$CO), and a median decrease of 0.3\% (0.4\% for $^{13}$CO). We list the total mass and area in clumps for each cloud in Table \ref{table:cloud_size}.  

\subsection{IR and Optical Star Formation Indicators} \label{sec:ir_optical_tracers}
We use several sets of data to estimate the star formation activity within our target molecular clouds:

\begin{enumerate}
\item $8\micron$ band observations from the SAGE IRAC Spitzer survey \citep{2006AJ....132.2268M, https://doi.org/10.26131/irsa404}, with 1\farcs 2 pixels. We match the point source subtracted $8\micron$ mosaic to 3\farcs 5 using the kernels provided in \cite{aniano_common-resolution_2011}.
\item Continuum-subtracted H$\alpha$ data with 5\arcsec\ resolution from the Magellanic Clouds Emission-Line Survey (MCELS; \citealp{smith_umctio_1998, gordonthesis2018}). H$\alpha$ traces massive, and hence the earliest ($\sim$ 10 Myr) stages of, star formation \citep{kennicutt_star_2012}. \cite{lopez_role_2014} outline the correction for extinction of H$\alpha$ using the \cite{haschke_new_2011} reddening maps of the LMC. The typical resolution of the reddening map is 4\farcm5. \cite{lopez_role_2014} find extinction corrections between 10 and 20\%. These coarse reddening maps may not be representative of the extinction in GMCs, so we do not employ this method in our study and note that the SFR may be underestimated by ignoring extinction effects.
\item We use the modified blackbody model in \cite{utomo_resolved_2019}, derived from \textit{Herschel Space Observatory} observations, for dust temperature ($T_d$). The map has 13 pc resolution, and, following \cite{chiang_spatially_2018}, uses a fixed value of emissivity ($\beta=1.8$) to break the degeneracy between $T_{d}$ and $\beta$ in the model. We also examine the \cite{gordon_dust_2014} modified blackbody model, using the same data but with $\beta$ fit independently for each pixel. Both \cite{gordon_dust_2014} and \cite{utomo_resolved_2019} characterize uncertainties on model fits by creating new model realizations after adding noise (20 realizations for the former, 100 for the latter). We find a nearly linear relationship between $T_d$ from the two models in our sample, so we proceed using the $T_d$ from \cite{utomo_resolved_2019}. Following \cite{draine_infrared_2007}, the energy density of the radiation field heating the dust, assuming the dust grains are in thermal equilibrium with the radiation field heating them (as in \citealp{utomo_resolved_2019}), is
\begin{equation} \label{eq:u_nu}
   u_{\nu} = 8.65\times 10^{-13} \left(\frac{T_d}{18\text{K}}\right)^{4+\beta} \text{ erg cm}^{-3}
\end{equation}
We adopt $u_{\nu}$ as our third tracer in preference to $T_d$ as it is more comparable to the other two tracers.
\end{enumerate}
Hereafter, we use the term ``star-formation tracers" to refer to the ensemble of $8\micron$, H$\alpha$, and $u_{\nu}$ data.

In Figure \ref{fig:N59C_RGB}, we show the H$\alpha$, 8$\micron$, and CO emission across N59C. While the H$\alpha$ and CO emission broadly do not trace each other, the CO-bright regions are often coincident with bright 8$\micron$ emission. The \textit{Spitzer} 8$\micron$ band is centered on a prominent PAH feature. Recent work (e.g., \citealp{whitcomb_star_2023}; \citealp{leroy_phangsjwst_2023}) has shown a strong correlation between CO and mid-infrared bands on sub-kpc and galactic scales. \cite{bendo_relations_2008} and \cite{cortzen_pahs_2019} have found the $8\micron$ and 6.2$\micron$ PAH bands to trace cold dust rather than warm dust. These results indicate that $8\micron$ could be preferentially tracing molecular gas column densities over star formation and the intensity of the radiation fields heating the dust and gas. Since the dust model from which $u_{\nu}$ is derived does not use $8\micron$ data, the three star-formation tracers we use are independent. Therefore, despite the coarseness of the pixels at which the model is sampled compared to the CO data, we use $u_{\nu}$ to sidestep uncertainties about the origin of the 8$\micron$ emission. Since H$\alpha$ directly traces ionizing stellar radiation \citep{kennicutt_star_2012} while $8\micron$ and $u_{\nu}$ trace the ionizing and non-ionizing radiation field heating grains (\citealp{li_infrared_2002, calzetti_calibration_2007}), the use of all three tracers can probe slightly different types of feedback. We compare results obtained with H$\alpha$ to those with $8\micron$ and $u_{\nu}$ to identify the relative strengths of different sources of feedback, if they are important in the energy budget of clumps.
\subsection{\texorpdfstring{$^{12}$CO-}{12CO-} and \texorpdfstring{$^{13}$CO}{13CO}-Based Clump Properties}
Using optically-thick $^{12}$CO, we can approximate a mass in structures by assuming a constant ratio between molecular hydrogen column density and $^{12}$CO brightness, called the $X_{CO}$ factor
\begin{equation} \label{eq:mlum}
    M_{\text{lum}} = 4.3 X_{CO} \frac{L_{CO}}{\text{K km s}^{-1} \text{pc}^2} M_{\odot}
\end{equation}
with $X(\text{J = 1 -- 0}) = 2.4$ \citep{hughes_physical_2010}, $X(\text{J = 2 -- 1}) = 3$ (adopting a constant line ratio $R_{21} = 0.8$ as in \citealp{wong_relations_2019}) for different transitions of CO. We have tested an alternative model in which $R_{21}$ is not constant to see if the amount of molecular gas inferred by CO emission varies locally. We used the statistical relationship derived from a survey of local galaxies at kpc scales between CO line ratio and 8$\micron$ presented in \cite{leroy_phangsjwst_2023},
\begin{equation}
    R_{21}\approx \text{min}\left(0.62 I_{8\micron}^{0.26}, 1\right)
\end{equation}
for $I_{8\micron}$ in $\text{MJy sr}^{-1}$. Using this prescription, $R_{21}$ for 30Dor, N159E, and N159W is 1, while $R_{21} \approx 0.582$ for PCC. These line ratios are consistent with prior measurements, with $R_{21} = 0.6 \pm 0.1$ in PCC \citep{wong_alma_2017} and $0.84 \pm 0.30$ in 30 Doradus and $1.22 \pm 0.11$ in N159 \citep{sorai_co_2001}. We find that as expected, using this prescription, the slope of the $\log$ \avgm  vs.\ $\log$ \avgS relation (further discussed in Section \ref{sec:caveats}) is steeper than in the case of uniform $R_{21}$, but the difference is not larger than the uncertainties associated with either fit. Because this prescription of $R_{21}$ does not significantly impact our results, we proceed with the analysis using the uniform value of $R_{21}$. While a cloud-by-cloud treatment of the low-J line ratio in the LMC does not yield results different from adopting a uniform value for the clouds in this sample, future studies using a larger sample of regions observed with different CO lines should evaluate the treatment of the line intensity ratio before interpreting results derived from CO brightness.

We follow the local thermodynamic equilibrium (LTE) analysis conducted in \cite{wong_alma_2017} and \cite{wong_relations_2019}. It is assumed that both the $^{12}$CO and $^{13}$CO lines for a given transition have the same excitation temperature, and that the excitation temperature $T_{ex}$ can be obtained from $^{12}$CO brightness. To compute mass from LTE calculations, we assume a constant $^{13}$CO to H$_2$ column density (from \citealp{wong_relations_2019} and references therein):
\begin{equation} \label{eq:abundance}
    \frac{N(\text{H}_2)}{N(^{13}\text{CO})} = 3 \times 10^6
\end{equation}
In general, this ratio is not necessarily constant, but can vary with isotopic abundance of $^{13}$C. Despite these assumptions, since $^{13}$CO is more optically thin, the LTE method should give more reliable measures of column density than that obtained from $^{12}$CO brightness and assuming a constant X$_{\text{CO}}$. We further discuss the caveats of interpreting LTE column densities in Section \ref{sec:caveats}. 

By assuming a structure is spherical and in virial equilibrium, a virial mass can be calculated 
\begin{equation} \label{eq:Mvir}
    M_{\text{vir}} = \frac{5 \sigma_v^2 R}{G}
\end{equation}
from the virial theorem. The virial parameter
\begin{equation} \label{eq:virial_parameter}
    \alpha = \frac{M_{\text{vir}}}{M_{\text{lum}}}
\end{equation}
is the ratio of virial and luminous masses and indicates the boundedness of clumps. For $^{13}$CO clumps, we can repeat these calculations, but replacing the luminous mass estimate with the LTE mass calculation. We also calculate the luminous surface density, $\Sigma_{lum}$ (LTE surface density $\Sigma_{LTE}$ for $^{13}$CO), and the virial surface density, $\Sigma_{vir}$, for each clump. In order to use consistent measures of clump sizes, these surface densities are the respective mass divided by the ``deconvolved" area, $\pi R_{d}^2$, using the deconvolved clump radius, rather than the exact clump area.

From the $^{13}$CO optical depths ($\tau$) produced during our LTE analysis, we calculate the average optical depth $\langle\tau_{13}\rangle$ for each clump. We find a few $^{12}$CO clumps with weak or undetected $^{13}$CO emission to have negative $\langle\tau_{13}\rangle$. Negative values of $\tau$ are not physical, but result from the linear treatment of the relationship between brightness temperature and $\tau$, as described in \cite{wong_relations_2019}. For these cases, we have investigated weighting $\tau_{13}$ by the CO signal-to-noise ratio (S/N), but this does not result in positive $\langle\tau_{13}\rangle$ for all of them. We decided to proceed by calculating $\langle \tau_{13} \rangle$ with equal weighting to all data rather than weighting by S/N.
\begin{figure*}[t] 
    \centering
    \includegraphics[width=\textwidth, trim={1.05in 0.45in 1.79in 1.045in},clip]{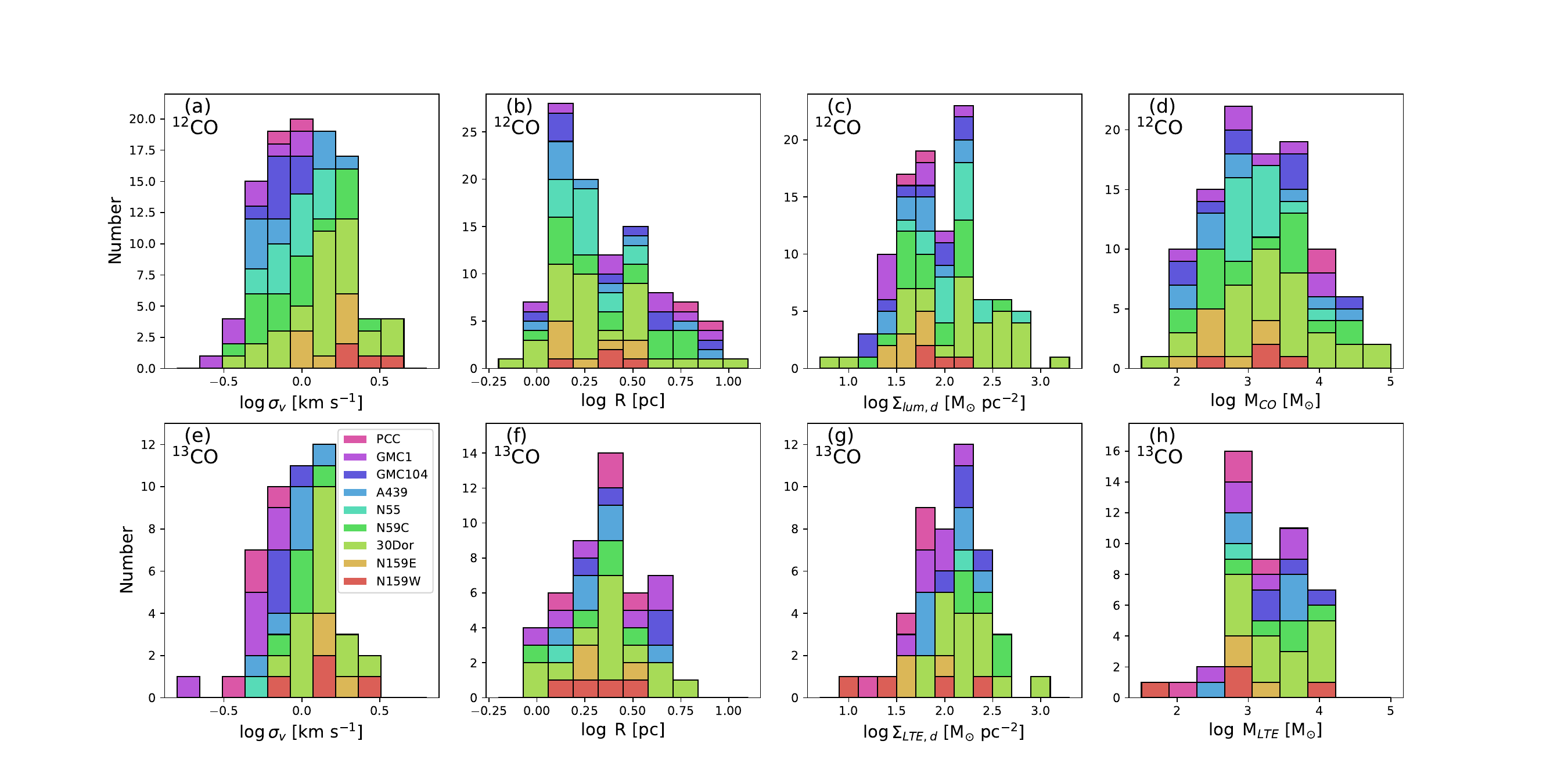}
    \caption{Histograms of clump (a, d) linewidths, (b, e) sizes, (c, f) surface densities, and  (d, h) CO-based (or LTE-based for $^{13}$CO) masses in  (a-d) $^{12}$CO and (e-h) $^{13}$CO. The linewidths and sizes presented are the deconvolved values, and the surface densities are calculated by dividing the mass by the deconvolved area, $\pi R_d^2$. The clumps identified range from sub-pc to 10 pc in size. Clumps are typically detected in $^{12}$CO at lower surface brightness than in $^{13}$CO. Clumps from the same cloud are assigned the same color (see also Figure \ref{fig:size_linewidth} for the color assignment scheme).}  
    \label{fig:histograms}
\end{figure*}
\subsection{Star-Formation Tracer-Based Clump Properties and Uncertainties} \label{sec:SF_tracer_treatment}
The IR/optical data are 2D maps, while the dendrogram structures are three-dimensional (position-position-velocity). To measure star-formation tracers in structures, we project the maximum extent of the dendrogram structure boundaries at any velocity channel onto the sky, and then calculate the average of each star-formation tracer map in this mask.

In order to estimate uncertainty in our clump-averaged measures of star formation indicators, we first estimate noise in the LMC mosaics from several signal-free regions of the maps. The regions used for noise-level estimates consist of 9.6\% of the native resolution pixels in the SAGE 8$\mu$m map, and $\sim$10\% of the pixels in the MCELS H$\alpha$ map. For 8$\mu$m, we convolve these cut-out regions to 3\farcs 5. We then compute the standard deviation from the robust MAD estimator. MAD is used to mitigate contamination from the bright point source-like objects in the fields. We then use the noise-addition method, described in Section \ref{sec:COsegmentation} on each dendrogram structure to compute an uncertainty on measured values of 8$\mu$m and H$\alpha$. Since pixels have correlated errors, we divide the uncertainty by the square root of the number of independent beams in the clump. We use the uncertainties computed on $T_d$ by \cite{utomo_resolved_2019} as the uncertainties on clump-averaged $T_d$. Because the emissivity $\beta$ is fixed in the \cite{utomo_resolved_2019} model, the fractional uncertainty in $u_{\nu}$ is simply the product of the fractional uncertainty in $T_d$ and the exponent $4 + \beta$ in Equation \ref{eq:u_nu}.
\section{Results} \label{sec:results}
\subsection{Clump properties}
\begin{figure*}[t] 
    {\centering \includegraphics[width=\textwidth, trim={2.49in 0.705in 3.195in 1.175in}, clip]{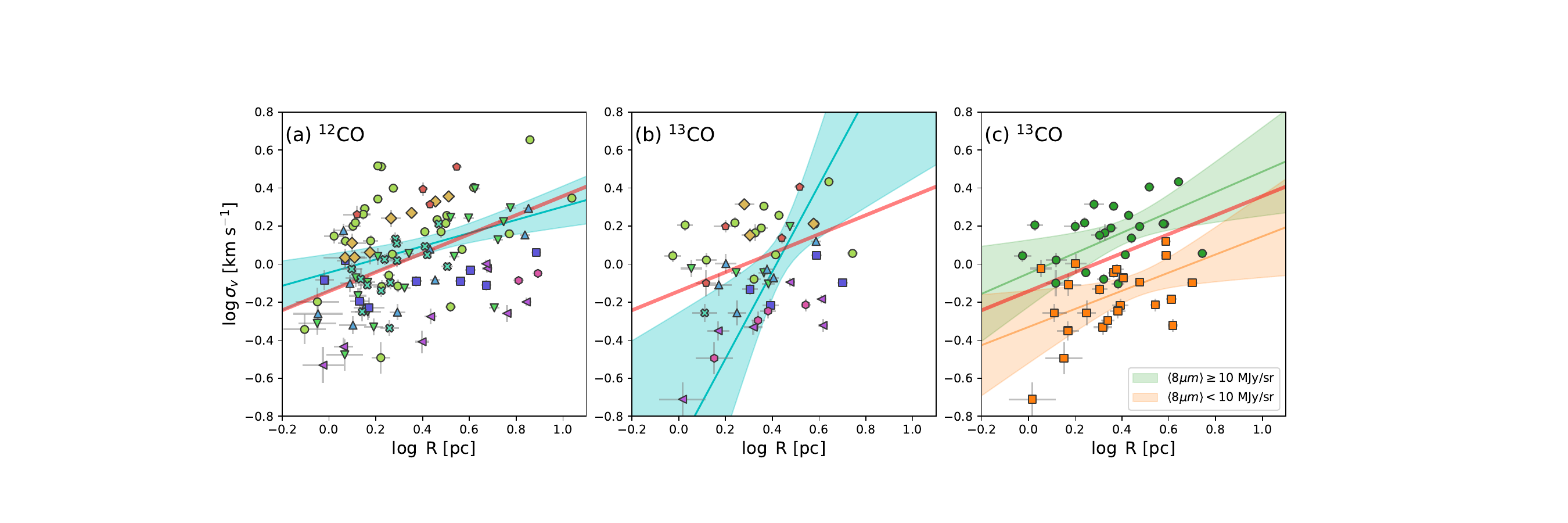}}
    \hspace*{0.45in} \includegraphics[height=0.6\textwidth, trim={0 0in 0 0}, clip, angle=270]{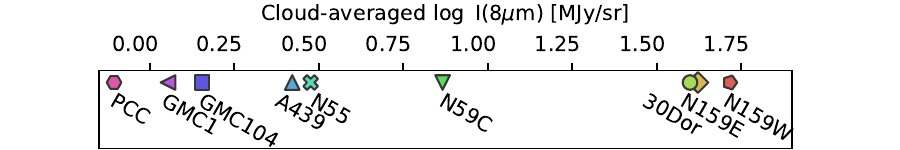}
        \newline
    \caption{The size vs.\ linewidth relations for (a) $^{12}$CO and (b) $^{13}$CO clumps. The fiducial \citetalias{solomon_mass_1987} relation is shown in red. The shaded cyan regions show 95\% confidence intervals of the $R-\sigma_v$ fit. Points are colored by the average 8$\mu$m value in the cloud, as shows in the color bar below panels (a) and (b). The $^{13}$CO clumps are reproduced in panel (c), but are colored by whether they have clump-averaged 8$\micron$ values above or below 10 MJy/sr. The ODR fits shown in this panel are fit only to the data in each of the $\langle 8\micron \rangle$ bins. The shaded regions again correspond to 95\% confidence intervals. The fit parameters and associated uncertainties for each model are presented in Section \ref{sec:SLW_fits}.}
    \label{fig:size_linewidth}
\end{figure*}

The distributions of clump linewidths, sizes, and surface densities are presented in Figure \ref{fig:histograms}. With sensitivity matched across the clouds, we probe surface densities down to about $10^{0.5} M_{\odot} \text{pc}^{-2}$. Several clumps are initially identified with radii below the beam size and linewidths smaller than the channel width due to the deconvolution methods described in Sections \ref{sec:resmatching} and \ref{sec:COsegmentation}, but few remain after applying the threshold on the size of the clump minor axis. Clumps identified in the $^{13}$CO data are similar to those in $^{12}$CO, except the latter can be detected at lower densities and larger scales. In both cases, the $\textit{SCIMES}$ clumps range from sub-pc to several pc sizes with linewidths typically between 0.3 -- 3 km s$^{-1}$.

We compute the mean column density of $^{13}$CO under the LTE assumption in each clump, ranging from $N\left(^{13}\text{CO}\right) = 10^{12.7} - 10^{15.7} \text{ cm}^{-2}$ in $^{12}$CO clumps and $N\left(^{13}\text{CO}\right) = 10^{14.2} - 10^{15.8} \text{ cm}^{-2}$ in $^{13}$CO clumps. As seen in the overlap of $^{12}$CO and $^{13}$CO clumps in Figure \ref{fig:scimes_overview}, $^{13}$CO is often only detected in the denser parts of CO-bright regions. This is evident from the lower column densities at which $^{12}$CO clumps are identified, and can be seen from areas in the maps with faint $^{12}$CO emission and no corresponding $^{13}$CO clumps. 

\begin{deluxetable*}{r|r|r|r|r|r|r}[htb!] \label{table:fiducial_relation}
    \tablecaption{Parameters, scatter, and correlation coefficient of fits to the residual of the fiducial relation. $m$ and $X_0$ are the parameters and uncertainties fit by ODR to the model $\log \nu_0 = m\left(X - X_0\right)$. $\varepsilon$ and $\varepsilon_{\text{norm}}$ are the vertical scatter and normalized vertical scatter, and $\rho$ is the Pearson correlation coefficient. The rows labeled ``$\log \nu_0$" note the scatter from the fiducial relation and correlation coefficient of the clump sizes and linewidths.}
    \tablehead{\colhead{} & \colhead{} & \colhead{$m$} & \colhead{X$_0$} & \colhead{$\varepsilon$} & \colhead{$\varepsilon_{\text{norm}}$} & \colhead{$\rho$}}
    \startdata
    \multirow{5}{*}{$^{12}$CO} & $\log \nu_0$ &  &  & 0.24 & 8.63 & 0.34 \\
    & $\log \nu_0$ vs.\ $\log$ \avgm & 0.27 $\pm$ 0.02 & 1.57 $\pm$ 0.06 & 0.16 & 4.79 & 0.74 \\
    & $\log \nu_0$ vs.\ $\log$ \avgH & 0.11 $\pm$ 0.02 & 4.87 $\pm$ 0.20 & 0.19 & 5.66 & 0.51 \\
    & $\log \nu_0$ vs.\ $\log$ \avgu & 0.41 $\pm$ 0.03 & -10.96 $\pm$ 0.05 & 0.19 & 5.71 & 0.66 \\
    & $\log \nu_0$ vs.\ $\log$ \avgS & 0.66 $\pm$ 0.06 & 2.25 $\pm$ 0.04 & 0.24 & 7.18 & 0.58 \\
    \hline
    \multirow{5}{*}{$^{13}$CO} & $\log \nu_0$ &  & & 0.23 & 7.26 & 0.29 \\
    & $\log \nu_0$ vs.\ $\log$ \avgm & 0.21 $\pm$ 0.02 & 1.98 $\pm$ 0.11 & 0.13 & 3.70 & 0.84\\
    & $\log \nu_0$ vs.\ $\log$ \avgH & 0.09 $\pm$ 0.01 & 5.30 $\pm$ 0.32 & 0.12 & 4.11 & 0.73\\
    & $\log \nu_0$ vs.\ $\log$ \avgu & 0.30 $\pm$ 0.04 & -10.72 $\pm$ 0.10 & 0.14 & 4.22 & 0.79\\
    & $\log \nu_0$ vs.\ $\log$ \avgS & 0.63 $\pm$ 0.12 & 2.40 $\pm$ 0.08 & 0.25 & 7.48 & 0.42
    \enddata
\end{deluxetable*}

\subsection{Size-linewidth relation fits} \label{sec:SLW_fits}
To fit the size-linewidth data, as well as all of the other linear fits presented in this paper, we perform orthogonal distance regression (ODR) using the $\textit{scipy.odr}$ package \citep{2020SciPy-NMeth}. For each set of $^{12}$CO and $^{13}$CO data, we fit the size-linewidth data of the whole multi-cloud population of clumps in order to identify deviations from the scaling relation, measure the physical state of molecular clouds, and correlate these deviations and physical properties with star formation activity. The best fit and the 95\% confidence intervals of the power-law size-linewidth relation are shown in Figure \ref{fig:size_linewidth} as a solid cyan line and shaded cyan region. We compare the fits to the size-linewidth relationship from \citetalias{solomon_mass_1987}, $\sigma_v = 0.72 \left(R/\text{pc}\right)^{0.5} \text{ km s}^{-1}$, hereafter the ``fiducial relation" (solid red line in Figure \ref{fig:size_linewidth}). The fit to the $^{12}$CO data, 
\begin{equation*}
    \log \left(\frac{\sigma_v}{\text{km s}^{-1}}\right) = \left( 0.35\pm 0.09\right) \log \left(\frac{R}{\text{pc}}\right) -  0.04\pm 0.05,
\end{equation*}
has slope consistent with the fiducial relation to 1.7$\sigma$, but the fit to the $^{13}$CO data,
\begin{equation*}
    \log \left(\frac{\sigma_v}{\text{km s}^{-1}}\right) = \left( 2.29\pm 0.78\right) \log \left(\frac{R}{\text{pc}}\right) - 0.95 \pm 0.35,
\end{equation*}
is much steeper than the fiducial relation (2.3$\sigma$ away from the fiducial slope), and the fit parameters have large relative uncertainties. The $^{12}$CO fit is well-constrained enough to reject a zero-slope relation at the $3.9\sigma$ level, but the $^{13}$CO fit slope is poorly constrained, with a slope of zero consistent at the $2.9\sigma$ level.

Motivated by the presence of several small, low-linewidth $^{13}$CO clumps, we test whether censoring these data result in a fit closer to the fiducial relation. However, due to the large uncertainties associated with these data leading to them being downweighted in the ODR fitting, censoring the small radius clumps neither brings the slope of the fit significantly closer to 0.5 nor reduces the size of the confidence interval. Instead, we find that fitting high- and low-$8\micron$ clumps separately does result in fits with similar slopes to the fiducial relation (panel c of Figure \ref{fig:size_linewidth}). For $^{13}$CO clumps with $I(8\micron) < 10 $ MJy sr$^{-1}$, we find a fit of
\begin{equation*}
    \log \left(\frac{\sigma_v}{\text{km s}^{-1}}\right) = \left(0.48 \pm 0.19\right) \log \left(\frac{R}{\text{pc}}\right) - 0.33 \pm 0.09
\end{equation*}
and for those with $I(8\micron) \geq 10$ MJy sr$^{-1}$, we find a fit of 
\begin{equation*}
    \log \left(\frac{\sigma_v}{\text{km s}^{-1}}\right) = \left(0.54 \pm 0.19\right) \log \left(\frac{R}{\text{pc}}\right) - 0.05 \pm 0.08.
\end{equation*}

The intercepts of these fits correspond to linear normalizations of $10^{-0.33} \approx 0.47$ and $10^{-0.05} \approx 0.89$. Separated this way, these normalizations bound the 0.72 normalization of the fiducial relation. The slopes are $< 1 \sigma$ away from the 0.5 slope of the fiducial relation. This test suggests that despite the fit to the $^{13}$CO size vs.\ linewidth data being inconsistent with the fiducial relation, the data are well-described by the combination of the fiducial relation and a third parameter describing the offset of data from this line. 

The range in linewidths of around one dex for clumps of a given size is consistent with previous observations of the size-linewidth relation across all scales at which it has been observed (see the review by \citealp{hennebelle_turbulent_2012}). In the first and sixth rows of Table \ref{table:fiducial_relation}, we present the RMS scatter, $\varepsilon$, from the fiducial relation, computed using the vertical distance from the data to the fit line. We also present the scatter normalized by the uncertainty, $\varepsilon_{\text{norm}}$, computed as the RMS of the vertical distances of data from the fit normalized by the uncertainty in the y-axis data. For example, $^{12}$CO clump linewidths are typically $8.63\sigma$ away from the fiducial relation, showing that the data do not just have large scatter about the fiducial relation ($\varepsilon = 0.24$ dex), but the scatter is also large relative to the uncertainty in the data. The Pearson correlation coefficient, $\rho$, of the $^{12}$CO size vs.\ linewidth data is 0.34, and 0.29 for the $^{13}$CO clumps, both indicative of only weakly correlated data.

\subsection{Fitting star-formation tracers and gas properties to the scatter in the size-linewidth relation} \label{sec:sft_correlations}
\begin{figure*}[!htb] 
    \centering
    \includegraphics[width=\textwidth, trim={2.7in 0.45in 3.19in 1.045in},clip]{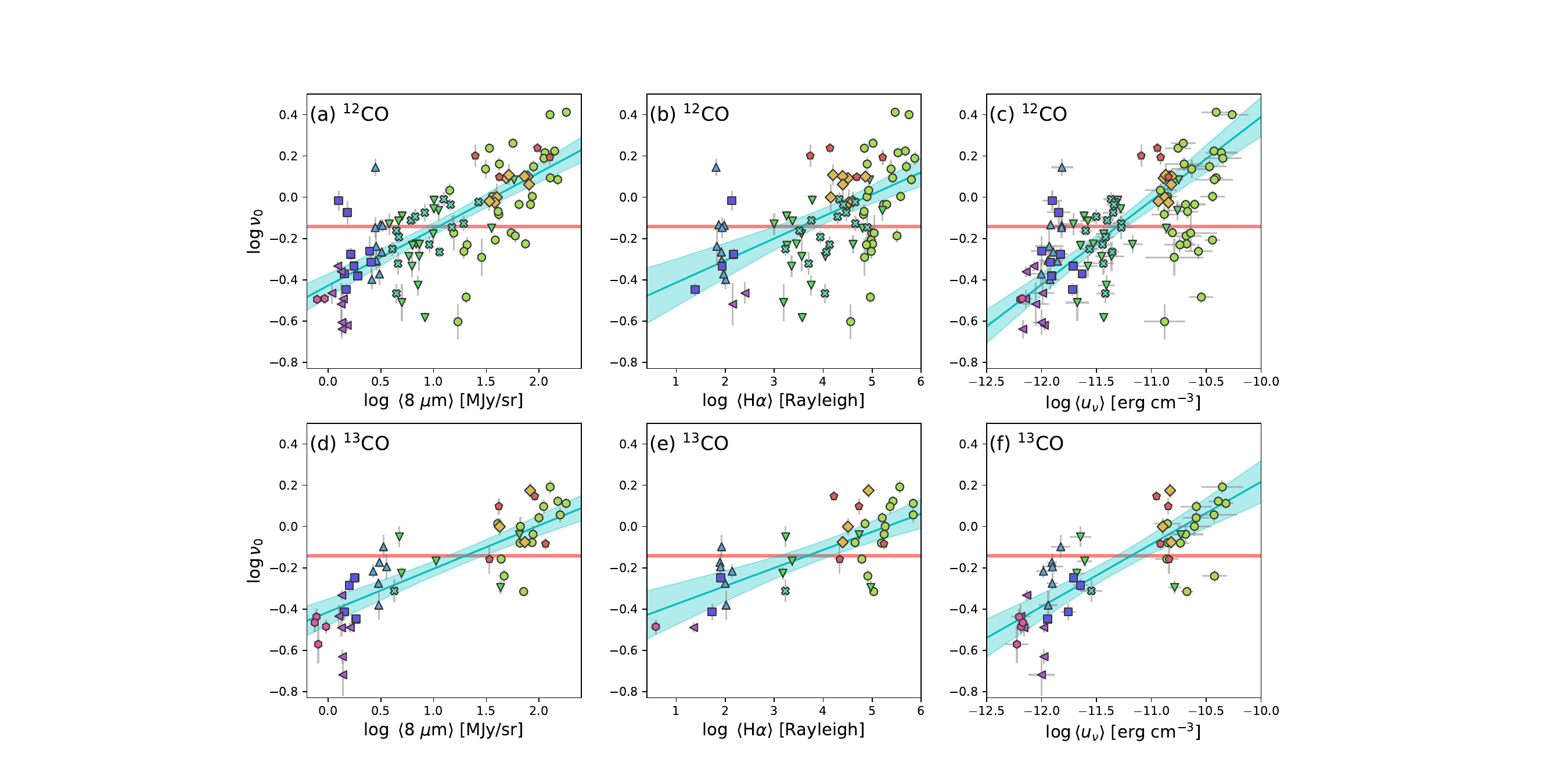}
    \caption{Star-formation tracers vs.\ $\log \nu_0$ for (a-c) $^{12}$CO and (d-f) $^{13}$CO clumps. Points are colored and have the same plot symbols as the color bar in Figure \ref{fig:size_linewidth}. The red horizontal line is plotted at $\nu_0 = 0.72$, the normalization of the fiducial relation. Fourteen of the 103 $^{12}$CO clumps and 10 of the 47 $^{13}$CO clumps are not included in the \avgH vs.\ $\nu_0$ fits due to bad or missing data. $\nu_0$ is correlated with each star-formation tracer, and the fits for each tracer is consistent between the $^{12}$CO and $^{13}$CO clump samples, though there is less scatter and higher correlation in the $^{13}$CO data.}
    \label{fig:nu0}
\end{figure*}
\begin{figure*}[!htb] 
    \centering
    \includegraphics[width=\textwidth, trim={1.41in 0.625in 2.115in 1.155in}, clip]{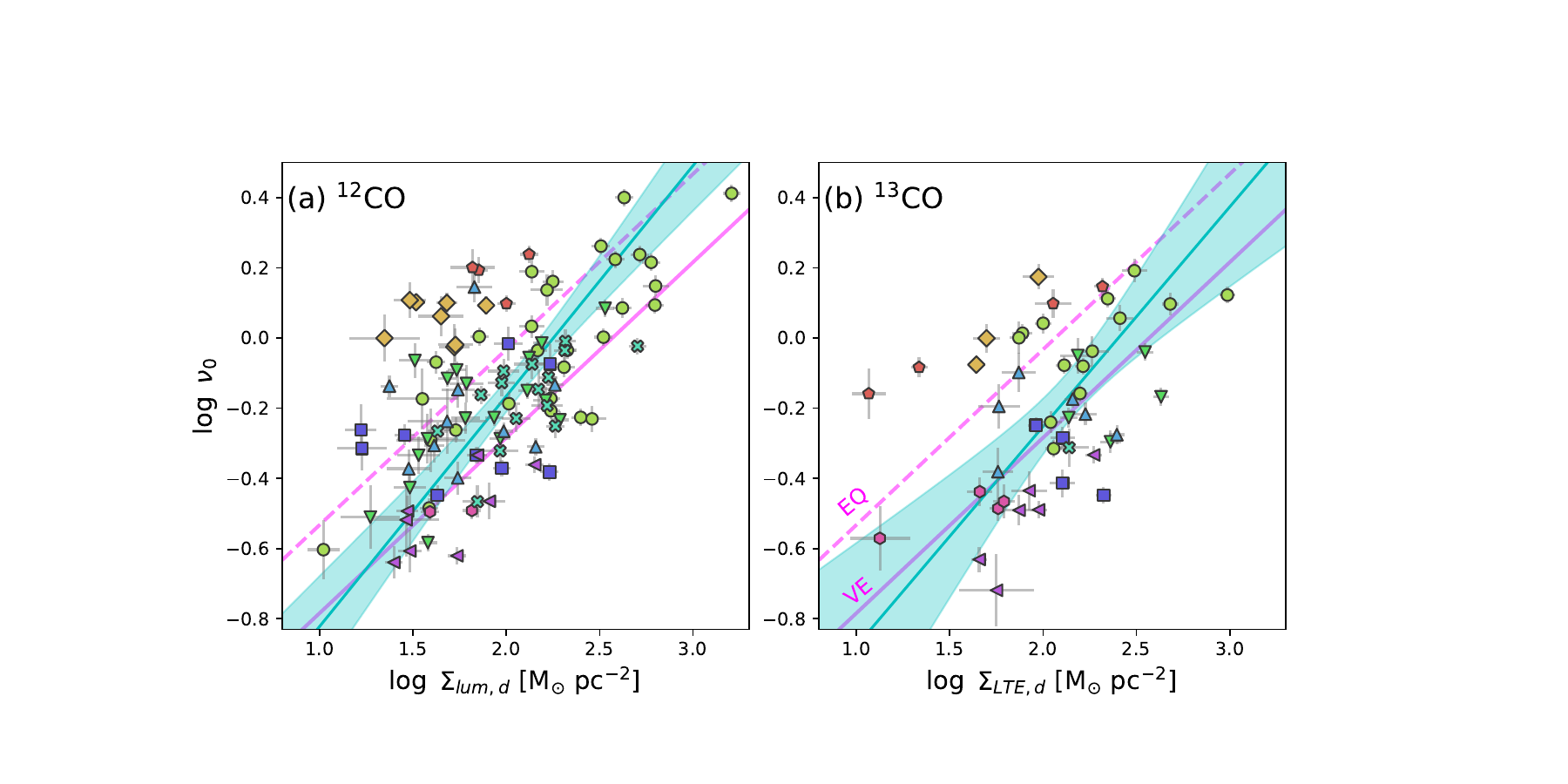}
    \caption{Surface density vs. $\log \nu_0$ for (a) $^{12}$CO and (b) $^{13}$CO clumps. The cyan line and shaded region show the ODR fit and 95\% confidence interval. The solid magenta line shows the condition of virial equilibrium for a uniform, spherical cloud ($\nu_0 = \sqrt{\pi G \Sigma / 5}$; \citealp{heyer_re-examining_2009}), while the dashed magenta line shows the condition of equipartition ($\nu_0 = \sqrt{2G\Sigma}$; \citealp{ballesteros-paredes_gravity_2011}).}
    \label{fig:nu0_sigma}
\end{figure*}
We fit the average star-formation tracers and surface densities of the multi-cloud population of clumps to the residual of the fiducial size-linewidth relation, $\log \nu_0$. These fits are shown in Figures \ref{fig:nu0} and \ref{fig:nu0_sigma}. The data are fit to models of the form 
\begin{equation}
    \log \nu_0 = m \left(X - X_0\right)
\end{equation}
given data $X$ and $\log \nu_0$ and model parameters $m$ and $X_0$. These fit parameters and associated uncertainties are presented in Table \ref{table:fiducial_relation}, alongside the goodness-of-fit metrics, $\varepsilon$, $\varepsilon_{\text{norm}}$, and $\rho$, described in Section \ref{sec:SLW_fits}.  In Appendix \ref{appendix:correlation} we present equivalent results when comparing the star-formation tracers with residuals from the best-fit, rather than fiducial, size-linewidth relation.  

Figure \ref{fig:nu0} reveals the three star-formation tracers show strong positive correlations with $\log \nu_0$ in both $^{12}$CO and $^{13}$CO clumps, with all six pairs of data having $\rho > 0.5$. In all of the star-formation tracer vs.\ $\log \nu_0$ fits, the slopes are similar between the $^{12}$CO and the $^{13}$CO data. The discrepancy is largest for the $\log \langle u_{\nu}\rangle$ vs.\ $\log\nu_{0}$ and the $\log \langle 8\micron\rangle$ vs.\ $\log\nu_{0}$ fits, in which the slopes of the fits for the two isotopologues are separated by $\sim 3\sigma$, while the separation is $\sim 2 \sigma$ for the $\log$ \avgH vs.\ $\log \nu_0$ fits. The $^{13}$CO data have lower $\varepsilon$ and $\varepsilon_{\text{norm}}$, as well as higher $\rho$ than the corresponding $^{12}$CO fits. Further, all six of these fits have higher $\rho$ than the size vs.\ linewidth data and lower $\varepsilon$ and $\varepsilon_{\text{norm}}$ than the fiducial relation by itself. The significant correlation coefficients and the smaller scatter in the star-formation tracer relations with $\nu_0$ demonstrates the ability of star-formation tracers to trace the excess linewidth from the fiducial size-linewidth relation. 

\begin{figure*}[t]
    \centering
    \includegraphics[width=\textwidth, trim={2.7in 0.45in 3.19in 1.045in},clip]{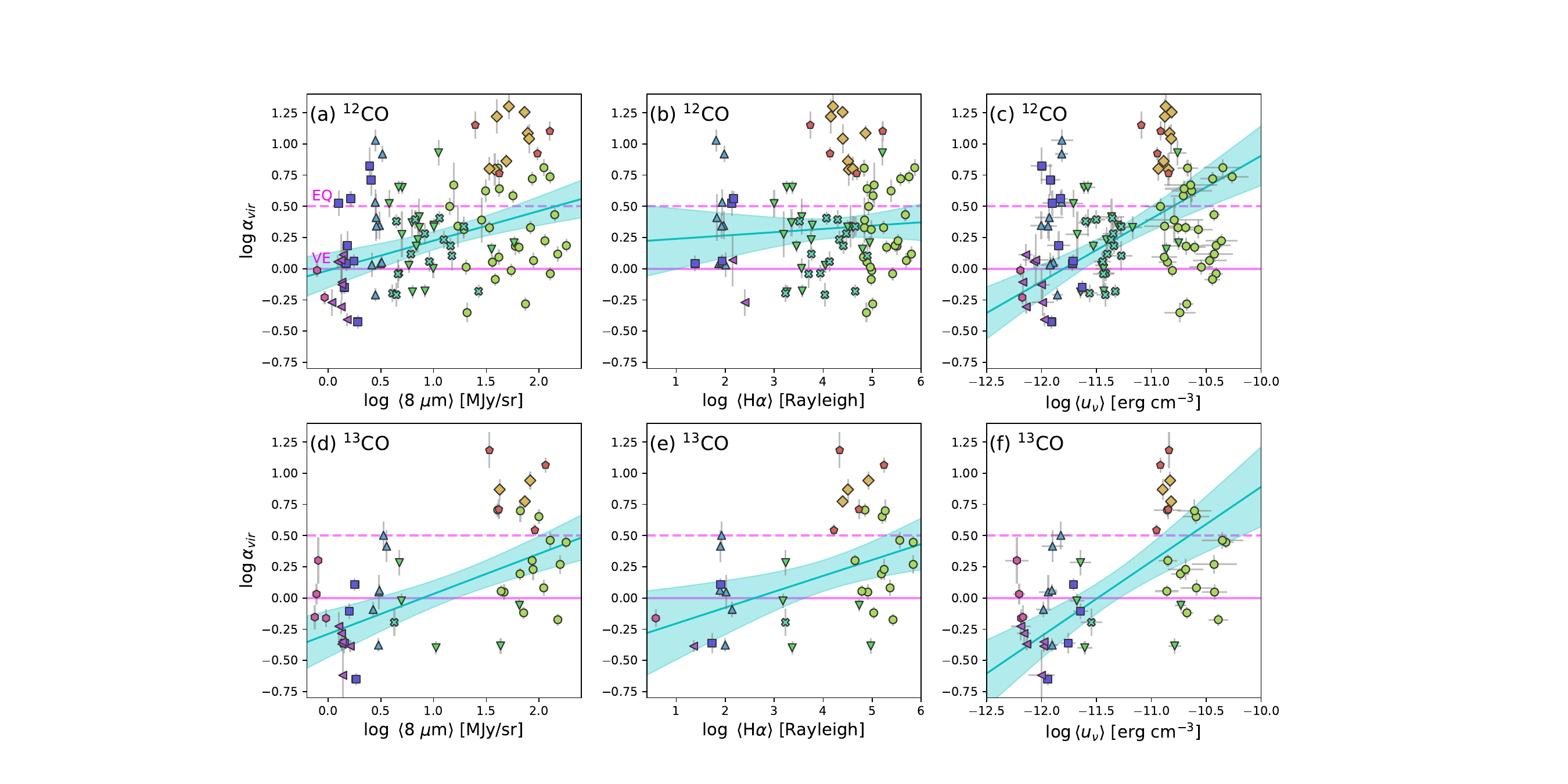}  
    \caption{Star-formation tracer vs.\ virial parameter for (a-c) $^{12}$CO and (d-f) $^{13}$CO clumps. The cyan line and shaded region show the ODR fit and 95\% confidence interval. The horizontal lines show the virial parameter for different cloud models: the solid magenta line shows a uniform, spherical cloud ($\alpha_{vir} = 1$) and the dashed magenta line shows a cloud in equipartition ($\alpha_{vir} = 10/\pi$). These panels show weak correlation with $\rho$ ranging from 0.32 to 0.58, except the \avgH vs.\ $\alpha_{vir}$ relation which is uncorrelated ($\rho = 0.09$).}  
    \label{fig:virial_parameter}
\end{figure*}
\begin{deluxetable*}{r|r|r|r@{ }c@{ }r|r|r|r}[hbt!] \label{table:virial_param_table}
    \tablecaption{Parameters, scatter, and correlation coefficient of ODR fits to virial parameter with  the model $\log \alpha_{\text{vir}} = m\left(X - X_0\right)$. The columns follow the conventions noted in Table \ref{table:fiducial_relation}.}
    \tablehead{\colhead{} & \colhead{} & \colhead{$m$} & \colhead{} & \colhead{X$_0$} & \colhead{} & \colhead{$\varepsilon$} & \colhead{$\varepsilon_{\text{norm}}$} & \colhead{$\rho$}}
    \startdata
    \multirow{3}{*}{$^{12}$CO} & $\log \alpha_{\text{vir}}$ vs.\ $\log$ \avgm & 0.24 $\pm$ 0.05 & 0.06 & $\pm$ & 0.28 & 0.38 & 5.42 & 0.39 \\
    & $\log \alpha_{\text{vir}}$ vs.\ $\log$ \avgH & 0.03 $\pm$ 0.04 & -8.02 & $\pm$ & 16.48 & 0.39 & 5.59 & 0.09 \\
    & $\log \alpha_{\text{vir}}$ vs.\ $\log$ \avgu & 0.50 $\pm$ 0.08 & -11.80 & $\pm$ & 0.11 & 0.42 & 6.00 & 0.32 \\
    \hline
    \multirow{3}{*}{$^{13}$CO} & $\log \alpha_{\text{vir}}$ vs.\ $\log$ \avgm & 0.32 $\pm$ 0.06 & 0.90 & $\pm$ & 0.17 & 0.38 & 5.39 & 0.58 \\
    & $\log \alpha_{\text{vir}}$ vs.\ $\log$ \avgH & 0.13 $\pm$ 0.04 & 2.61 & $\pm$ & 0.68 & 0.40 & 5.95 & 0.42 \\
    & $\log \alpha_{\text{vir}}$ vs.\ $\log$ \avgu & 0.60 $\pm$ 0.11 & -11.49 & $\pm$ & 0.10 & 0.41 & 6.14 & 0.53 \\
    \enddata
\end{deluxetable*}
In $^{12}$CO clumps, \avgm is the star-formation tracer that best explains the variance in $\nu_0$, given that this relation has the highest $\rho$, lowest $\varepsilon$, and lowest $\varepsilon_{\text{norm}}$ of the set of three tracers. In $^{13}$CO clumps, \avgm similarly has the highest $\rho$ and lowest $\varepsilon_{\text{norm}}$, but the \avgH vs.\ $\nu_0$ fit has slightly lower $\varepsilon$. Because the $\log$ \avgm vs.\ $\log \nu_0$ fit has lower $\varepsilon_{\text{norm}}$ but higher $\varepsilon$ than the $\log$ \avgH vs.\ $\log \nu_0$ fit, the outliers from the former fitted relation are typically those which are less robustly constrained in either size, linewidth, or a combination of the two (e.g., clumps with larger uncertainties arising from deconvolution or clumps with non-Gaussian line profiles). The fits involving $\log$ \avgu have comparable goodness-of-fit metrics as the other star-formation tracers, though never the best of the set. Because the resolution of the dust model used is coarse relative to the size of typical clumps, the fractional uncertainty in $\log$ \avgu is often much higher than the uncertainties in the other star-formation tracers (large horizontal error bars in panels (c) and (f) of Figure \ref{fig:nu0}).

Despite the strong correlations found between $\log \nu_0$ and star-formation tracers, one needs to ask if the star-formation tracers are the causal factor for the offset from the fiducial relation or if both are merely correlated with another factor. The most likely other causal factor is the surface density, $\Sigma$, since surface density and $\nu_0^2$ are proportional if clumps are in self-gravitational equilibrium \citep{heyer_re-examining_2009}. We show the \avgS vs.\ $\nu_0$ relations in Figure \ref{fig:nu0_sigma}. While the $^{12}$CO \avgS vs.\ $\nu_0$ data show stronger correlation than the $^{12}$CO \avgH vs.\ $\nu_0$ data, it has larger $\varepsilon$ and $\varepsilon_{\text{norm}}$ than any of the three star-formation tracer relations. The $^{13}$CO \avgS vs.\ $\nu_0$ relation also has weaker correlation and higher $\varepsilon$ and $\varepsilon_{\text{norm}}$ than all of the star-formation tracer relations. Compared to the star-formation tracers, the $^{12}$CO \avgS vs.\ $\nu_0$ relation shows only modest improvement over the fiducial relation in uncertainty-scaled scatter, and no significant improvement in the absolute scatter. On the other hand, the $^{13}$CO \avgS vs.\ $\nu_0$ relation has both larger absolute and uncertainty-scaled scatter than the simple fiducial relation. This is possible because the ODR fits weight data by the inverse square of the uncertainty for both the independent and dependent variables, so data with larger uncertainty contribute less to the fitted model, but $\varepsilon$ is an unweighted RMS, and $\varepsilon_{\text{norm}}$ is weighted only by the uncertainty in $\nu_0$.

Because the $\log$ \avgS vs.\ $\log \nu_0$ data are consistent with a 0.5 power-law slope, models in which clump linewidths are driven by gravity (whether clumps are collapsing or maintaining virial equilibrium) can explain some of the variance in the size-linewidth relation at the observed spatial scales. However, we have found that 1.) the surface density has lower correlation coefficient with $\nu_0$ than all of the star-formation tracers except for the \avgH data for $^{12}$CO clumps, and 2.) the surface density fits to the $\nu_0$ data has higher scatter (measured through $\varepsilon$ and $\varepsilon_{\text{norm}}$) than the star-formation tracer fits. Therefore, the star-formation tracers explain more of the variance in the size vs.\ linewidth data than the surface density. As well, the $\log$ \avgS vs.\ $\log \nu_0$ data do not seem to be well explained by a single normalization. The predictions of models without stellar feedback compared to the $\log$ \avgS vs.\ $\log \nu_0$ data for these clumps is discussed further in Section \ref{sec:driving_excess_linewidths}.

\subsection{Correlations between star-formation tracers and virial parameter}
The virial parameter (Equation \ref{eq:virial_parameter}) is proportional to $\nu_0^2$ and inversely proportional to surface density by definition, so lines of constant virial parameter increase diagonally to the upper left in the $\Sigma$ vs.\ $\nu_0$ plots (Figure \ref{fig:nu0_sigma}). To investigate how much of the scatter in the $\Sigma$ vs.\ $\nu_0$ relation star-formation tracers can explain, we plot star-formation tracers against $\alpha_{vir}$ in Figure \ref{fig:virial_parameter}  and present fit parameters in Table \ref{table:virial_param_table}. The predictions of clumps in virial equilibrium and equipartition are shows as lines of constant $\alpha_{vir}$. 
The virial parameters of spheres confined by external pressure can be calculated from the results in \cite{field_does_2011}, and they lie between the values of virial parameter for equipartition and virial equilibrium. The lines of constant virial parameter in Figure \ref{fig:nu0_sigma} then bracket the range of expected virial parameters for spherical, non-magnetic clumps from gravity-driven models.

Except for the $^{12}$CO $\log$ \avgH vs.\ $\log \alpha_{vir}$ relation which shows no correlation, the rest of the star-formation tracers have weak but positive correlation with virial parameter ($0.3 < \rho < 0.6$). However, even in the aforementioned case of no linear correlation, a single value of virial parameter does not fit the ensemble of clumps well. Further, many clumps have virial parameter far in excess of the predicted virial parameter of a clump in free-fall collapse, with some having $\alpha_{vir} > 10$. While the weak correlation coefficients in these relations preclude these star-formation tracers from being used to unambiguously explain the residuals in the gravity-driven $\nu_0 \propto \Sigma^{1/2}$ models, the fact that clumps' virial parameters do not have similar values and that some have large values of $\alpha_{vir}$ demonstrate that there are observations of clump energies which these models have yet to explain.

\begin{figure}[t]
    \centering
    \includegraphics[width=0.48\textwidth, trim={0 0 0.475in 0.5in},clip]{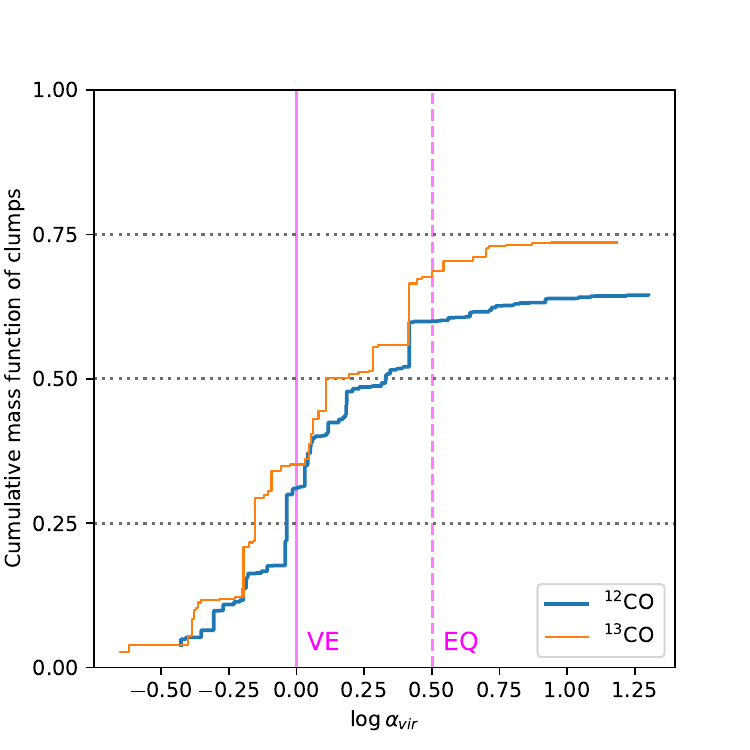}
    \caption{Cumulative mass function of (blue) $^{12}$CO and (orange) $^{13}$CO clumps vs.\ virial parameter. The virial parameters corresponding to virial equilibrium and equipartition are labeled ``VE" and ``EQ" next to solid and dashed magneta lines, respectively. The high-$\alpha_{vir}$ tails of the curves end below mass fraction of 1 because not all of the masked emission in the cubes is bound by clumps in our sample. The total mass traced by $^{12}$CO is found using Equation \ref{eq:mlum} for every pixel in the moment mask for each $^{12}$CO cube. The total mass traced by $^{13}$CO is found from the LTE-derived column density maps and the assumed constant $^{13}$CO-to-H$_2$ abundance (Equation \ref{eq:abundance}).}
    \label{fig:mass_frac}
\end{figure}
\section{Discussion} \label{sec:disc}
The standard interpretation of the size-linewidth relationship is that it represents the cascade over many orders of magnitude in scale of turbulent energy from injection into molecular clouds \citep{chevance_life_2023}. A tighter size-linewidth-surface density relation has been shown to exist in GMC observations, with this latter relation implying that molecular clouds all have the same ratio of kinetic and gravitational potential energy. Depending on the normalization of the relation, models which satisfy the size-linewidth-surface density relation explain the non-thermal linewidths as arising from virial balance or hierarchical collapse. In either case, the kinetic energy of clumps is gravity-driven. We have found that parsec-scale clumps in a resolution- and noise-matched sample of molecular clouds across a diverse range of star forming conditions span an order of magnitude in virial parameter. We find that the mean of our data do not match the expectation for either the virial equilibrium or the equipartition scenario of gravity-driven linewidths. We also find linear correlations between several independent star-formation tracers and the offset from the mean size-linewidth relation, as well as weaker correlations between star-formation tracers and virial parameter. 
\subsection{What is responsible for driving excess linewidths in parsec-scale emission?} \label{sec:driving_excess_linewidths}
Agreeing with previous observational work, we find a size-linewidth-surface density relationship to fit our data well, but while some of the $^{12}$CO clumps are consistent with the picture of energy equipartition turbulent collapse  given by \cite{ballesteros-paredes_gravity_2011}, $\nu_0 = \sqrt{2G\Sigma}$, many of our clumps are in excess of this condition (Figure \ref{fig:nu0_sigma}).

From the three goodness-of-fit metrics presented in Table \ref{table:fiducial_relation}, all three of the star-formation tracers considered explain more of the variance in the size-linewidth relation than surface density and show much stronger linear relationships with $\log \nu_0$ than those of the surface densities. The square of the Pearson correlation coefficient indicates the amount of variance in the data explained by the independent variable. While the heteroskedasticity of these data make it difficult to claim that some fraction of the variance in the data is definitely explained by a particular tracer, this behavior indicates that clump linewidths respond to local star formation activity in a uniform way, whereas the clump surface density has weaker explanatory power for clump linewidths. In essence, a single gravity-driven model does not fully explain the linewidths of the ensemble of clumps. 

The three star-formation tracers we have considered do not trace exactly the same feedback physics. CO brightness and 8$\micron$ emission are correlated owing to the sensitivity of 8$\micron$ to gas surface density, but 8$\micron$ is also excited by the local radiation field. If $8\micron$ was purely tracing surface density, the \avgm vs.\ $\nu_0$ correlation would be expected to mirror the \avgS vs.\ $\nu_0$ correlation. However, the latter have slopes 2.5 and 3 times steeper than the former for $^{12}$CO and $^{13}$CO, respectively. Furthermore, \avgm and \avgS are not strongly correlated in our sample. In Appendix \ref{appendix:tracer_correlations}, we present the mutual correlations between all of the star-formation tracers and the surface density in clumps. Clump-averaged $8\micron$ values are linearly correlated with $u_{\nu}$, and are much more weakly correlated with surface densities, especially in $^{13}$CO, which is expected to more faithfully trace relative column density than $^{12}$CO. These evidence suggest that the correlation between the normalization of the size-linewidth relation and the star-formation tracers is not due to an underlying correlation between the star-formation tracers and gas surface density. Future investigations using independent probes of density and temperature measured within the $^{12}$CO structures would increase the confidence in the reliability of 8$\micron$ as a star-formation tracer.

Because $u_{\nu}$ is sensitive to both ionizing and non-ionizing radiation field, it is a good tracer of stellar feedback independent of the number of massive stars in a cluster. $u_{\nu}$ shows similar trends with $\nu_0$ and $\alpha_{vir}$ as $8\micron$, with the weaker correlation ($\rho = 0.32$ in $^{12}$CO and $0.53$ in $^{13}$CO) and larger scatter in the trends likely reflecting the coarse resolution of the dust temperature models relative to the CO data. 

H$\alpha$ is not well correlated with the parsec-scale CO emission (see the \avgH vs.\ \avgS correlations presented in Appendix \ref{appendix:tracer_correlations}), instead tracing the ionized gas at localized star-forming regions in GMCs.  The correlation between \avgH and virial parameter in $^{13}$CO is comparable to the correlations with the other tracers, but the same correlation is not exhibited in the $^{12}$CO data. Whether the difference between the correlation in the two lines is due to difference in the spatial extent of the gas traced by $^{12}$CO vs.\ $^{13}$CO, or related to properties of H$\alpha$ as a feedback tracer remains to be answered. Analyses that might provide stronger evidence for this scenario could include correlating H$\alpha$ with the non-thermal component of linewidth and its trend as a function of distance from the stellar cluster center, as in \cite{2019MNRAS.489.4771G}, and using [CII] as a probe of an intermediate stage of feedback between CO linewidths and H$\alpha$. 

Because three independent feedback tracers show very similar correlations in $\nu_0$, and supported by some evidence of correlations with $\alpha_{vir}$, these results hint that feedback is important for explaining offsets from the size-linewidth (or size-linewidth-surface density) relation, and have a role in increasing the kinetic energy of parsec-scale molecular clumps.

In order to gauge how strong of an effect the high virial parameter clumps have on the dynamics of clouds as a whole, we examine how much of the total mass of clouds this population constitutes. In Figure \ref{fig:mass_frac}, we show the cumulative fraction of the mass in clumps that have $\alpha_{vir}$ below a given value. We normalize the mass functions to the total mass traced by $^{12}$CO and $^{13}$CO brightness in the masks described in Section \ref{sec:resmatching}, rather than just the total mass in the clumps studied in this sample. For this reason, the cumulative mass function at the highest measured value of $\alpha_{vir}$ does not equal unity. Instead, the $^{12}$CO clumps contain 65\% of the total mass traced by $^{12}$CO and the $^{13}$CO clumps contain 74\% of the total mass traced by $^{13}$CO. Despite the cuts imposed on the initial selection of clumps, our decomposition of the clouds has good flux recovery. The missing flux is distributed between the clumps close to the beam size which had been cut out of the sample (Section \ref{sec:COsegmentation}) and the diffuse, extended flux present in the masked emission (which extend down to 2$\sigma$ contours, Section \ref{sec:resmatching}).

Although there is a large number of clumps that have virial parameter above equipartition (30\% of the $^{12}$CO clumps and 21\% of the $^{13}$CO clumps), this population only contains a small amount of the molecular gas mass. In $^{12}$CO, 31\% of the total mass of the 9 clouds is in clumps below virial equilibrium, 29\% is in clumps between virial equilibrium and equipartition, and 5\% is in clumps above equipartition. In $^{13}$CO, the mass fractions in the same virial parameter bins are 35\%, 33\%, and 6\%, respectively. If the scalings between star-formation tracers and virial parameter are indeed reflecting the injection of kinetic energy by feedback, feedback appears to be a dominant energy source only for a small amount of the gas. It is possible that because clumps which are proximate to high amounts of feedback are subject to intense radiation fields, there is more CO photodissociation in these clumps. Tracers of CO-dark molecular gas, like [CII], could be used to test this conjecture.  

There is more mass in clumps below virial equilibrium than in clumps between virial equilibrium and equipartition, which suggests that the simple treatment of virial equilibrium used in this analysis does not fully describe these clumps and other energy terms, such as contributions from magnetic fields, are important to understand the dynamics of these clumps. 

Our sample covers molecular clouds of diverse star formation conditions at matched sensitivity and resolution, and is representative of GMCs in the LMC. Because the correlations in the star-formation tracers vs.\ $\alpha_{vir}$ relationships are weak, it is unclear from the sample size and available data whether virial parameter and feedback energy do follow a relationship not yet apparent from these data, or whether the relationship is inherently scattered. In order to better understand the effects of feedback on molecular gas, and to disentangle the different feedback effects that contribute to the tracers that have been employed, individual sources could be studied to constrain the effects of certain types of feedback when others are absent. We discuss other limitations and caveats of this analysis in Section \ref{sec:caveats}.

\subsection{Is opacity broadening significant in this sample?} \label{disc:opacity}
\begin{figure}[t] 
    \includegraphics[width=0.48\textwidth, trim={0.33in 2.41in 1.045in 2.92in}, clip]{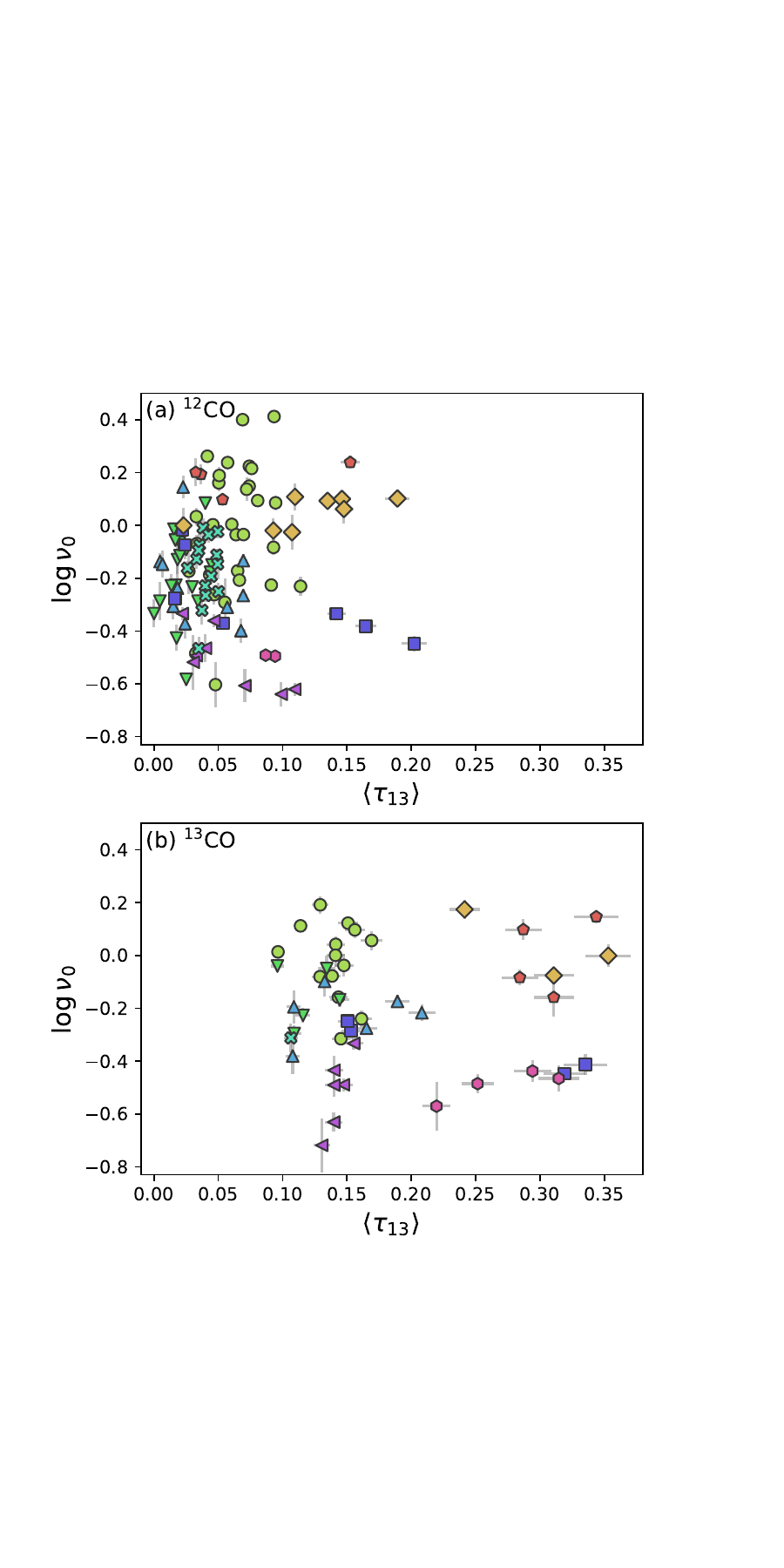}
    \caption{Clump-averaged optical depth of $^{13}$CO emission $\langle\tau_{13}\rangle$ vs $\log \nu_0$ for (a) $^{12}$CO clumps and (b) $^{13}$CO clumps. Optical depth is uncorrelated with the normalization of the size-linewidth relation. This suggests that optical depth effects, such as linewidth broadening in optically thick regions, has little effect on linewidth or $\nu_0$ trends observed across the entire sample of clouds.}
    \label{fig:tau}
\end{figure}
We infer the turbulent motions in molecular clouds from CO isotopologue line profiles. These emission lines are not just affected by bulk motions in the cloud however, as they are also subject to opacity effects. If 8$\micron$ is a stronger gas column density tracer than star formation activity tracer, the observed correlations between $8\micron$ and $\nu_0$ would be representing a more fundamental underlying column density vs.\ $\nu_0$ relation, with higher column density lines of sight being more optically thick. \cite{hacar_opacity_2016} describe that the low-J transtitions of the $^{12}$CO and $^{13}$CO isotopologues in optically-thick regions can be observed to have large linewidths due to opacity broadening or line blending. They note that in LTE conditions, optical depth of higher-J transitions of CO is lower, but the case is not as simple in non-LTE conditions. If LTE and abundances of CO isotopologues are assumed, the opacity determined from optically-thin C$^{18}$O can be scaled up to determine those of $^{12}$CO and $^{13}$CO for a line-of-sight. While observations of C$^{18}$O exist for all of the clouds in the sample except for N55, the line is detected very sparsely in each field and at low S/N.

We present the correlation between the average optical depth of the $^{13}$CO line, $\langle\tau_{13}\rangle$, and offset from the size-linewidth relations in Figure \ref{fig:tau}. Because it is often not well constrained, we adopt a uniform 5\% uncertainty on $\langle \tau_{13} \rangle$. In pixels with $^{12}$CO but low signal detections of $^{13}$CO, the estimates of optical depth are dominated by noise and can result in negative $\tau_{13}$. We do not censor these pixels while calculating clump averages, but we do not plot clumps with $\langle \tau_{13} \rangle < 0$. We find that $\langle \tau_{13} \rangle$ and $\nu_0$ are uncorrelated ($|\rho| < 0.2$). The fits are poorly described by a linear model, and have higher $\varepsilon$ than the simple $R-\sigma_v$ fits, suggesting that any linewidth broadening associated with higher optical depth does not account for the excess linewidth observed from the $R-\sigma_v$ relationship. Strictly, opacity broadening changes the shape of the emission line (broader, flat top) and causes line blending, so a more careful line fitting analysis would be required to correct for broadening, but this analysis would be performed on the beam scale rather than the clump scale. Additionally, $\langle \tau_{13} \rangle$ is only a measure of the optical depth of the $^{13}$CO line, and we calculate it in both $^{12}$CO and $^{13}$CO clumps. One of the assumptions in our LTE analysis is that $^{12}$CO is optically thick at the line center. We do not observe systematically larger linewidths in $^{12}$CO than in $^{13}$CO, which is predicted from opacity broadening (see Figure \ref{fig:histograms}), likely because the clumps we study are in the regime where linewidths are dominated by turbulence rather than opacity broadening. Again, this test should also be performed at the beam scale, as individual sightlines may better reflect any opacity broadening effects that are significant in these data. 

\subsection{Caveats and future directions} \label{sec:caveats}
Two major caveats of prescriptive use of the star-formation tracer vs.\ $\nu_0$ correlations arise that warrant discussion. First, the star-formation tracers have been averaged over the extent of the dendrogram boundaries, with typical radii of a few parsec. The variation in $8\micron$ emission around these parsec-scale CO peaks reflects the differences in dust heating by local variations in the interstellar radiation field. This is not directly comparable with studies which integrate over much larger ($\sim$ kpc) scales, in which $8\micron$ is found to trace star formation rate, as well as CO emission, with indications that $8\micron$ is a stronger CO tracer than a star formation rate tracer \citep{leroy_phangsjwst_2023, whitcomb_star_2023}. Because we adopt a CO clump-finding method, we are studying variations in $8\micron$ brightness among CO bright regions, rather than studying variations in $8\micron$ brightness across the ISM unbiased to the presence of molecular gas. We highlight this to note that the clump-averaged $8\micron$ is less sensitive to gas surface density, and to note that the use of $8\micron$ outside of clump-finding methods may be more sensitive to gas surface density and less sensitive to tracing star formation.

Second, the correlation between $8\micron$ and SFR is very sensitive to metallicity \citep{calzetti_calibration_2007}. The LMC has a shallower (\citealp{li_photometric_2023, choudhury_vmc_2021, grady_magellanic_2021, cioni_metallicity_2009}) metallicity gradient than the Milky Way, which also has evidence of a steeper gradient in the inner galaxy than the outer galaxy (\citealp{luck_distribution_2011, pedicelli_metallicity_2009}). The more uniform metallicity of the LMC suggests that our comparison of $8\micron$ between GMCs at different galactocentric radii may be appropriate, but caution should be applied when studying GMCs in galaxies with steep metallicity gradients, or when comparing GMCs from different galaxies. For robustness against metallicity, rather than using $8\micron$ to proxy the star formation activity for comparison between regions in different galaxies, mid-infrared tracers of feedback should be used to obtain relevant quantification of feedback type and strength.

We also note systematic uncertainties of our mass estimates. In addition to the optical depth of $^{12}$CO, another assumption of our LTE method is that $^{12}$CO and $^{13}$CO have the same excitation temperature. \cite{finn_physical_2021} analyze the Quiescent Molecular Ridge (south of the N159 regions studied in this work) with the non-LTE \texttt{RADEX} model \citep{van_der_tak_computer_2007}, and compare the derived molecular gas masses of clumps to those found assuming LTE. They found that, in comparison to $\texttt{RADEX}$, LTE underestimates mass for the J = 2 -- 1 lines, which they suggest could be due to subthermally excited $^{13}$CO, breaking the aforementioned LTE assumption. They also find overestimated LTE masses from the 1 -- 0 line, which could be due to optically thin $^{12}$CO. These results demonstrate that non-LTE conditions exist in the LMC.

LTE-based masses diverge from true mass at low $^{13}$CO column density, $\lesssim 10^{14}$ cm$^{-2}$, and gas volume density, $\lesssim 10^3$ cm$^{-3}$ (see \citealp{indebetouw_alma_2013} and references therein). While we do not have volume density estimates, 10\% of the $^{12}$CO clumps have mean column densities below $10^{14}$ cm$^{-2}$ and all of the $^{13}$CO clumps have higher average column densities than this bound. Since we do not use LTE-based masses to characterize the $^{12}$CO clumps, we do not expect the presence of subthermally excited $^{13}$CO in these regions to affect our results, but the effects of optically thin $^{12}$CO on our mass estimates remains uncertain. 

Finally, the results presented in this work are limited in sensitivity to gas surface densities of several $M_{\odot} \text{ pc}^{-2}$. It could be expected that feedback-driven turbulence is more efficient in clumps with less self-gravity, but observing this will require deeper observations with higher dynamic range.
\section{Conclusions} \label{sec:concl}
We have studied the size-linewidth relation and its residuals in resolution- and noise-matched LMC molecular cloud observations. We find that star-formation tracers explain more of the variance in the residuals of the size-linewidth relation than the molecular surface density can due to higher correlation coefficients and lower RMS scatter. We find that clumps do not scatter around a single value of virial parameter, as predicted by several models in which turbulent motions are purely gravity-driven. However, there is some evidence that the virial parameter is correlated with the local star formation activity. We do not find evidence that the observed offsets in linewidth from the size-linewidth relation can be explained by broadened linewidths from optically-thick regions. These results suggest that at parsec scales, feedback may contribute dynamically-significant amounts of energy or momentum to molecular clumps. However,  we still find the coefficient of the size-linewidth relation approximately scales with the square root of the cloud surface density in our data ($\nu_0 \propto \Sigma^{1/2}$),  suggesting that models predicting the virial equilibrium or energy equipartition of clouds are still important in understanding the dynamical state of molecular gas. Future models and observations should be able to disentangle the contributions to the turbulence in clumps from gravity and from feedback to fully explain the large scatter about the size-linewidth relation in observations of molecular clouds. 

\begin{acknowledgments}
This paper makes use of the following ALMA data: ADS/JAO.ALMA \#2012.1.00554.S, ADS/JAO.ALMA \#2013.1.00214.S, ADS/JAO.ALMA \#2013.1.00832.S, ADS/JAO.ALMA \#2016.1.00193.S, ADS/JAO.ALMA \#2019.1.00843.S. ALMA is a partnership of ESO (representing its member states), NSF (USA) and NINS (Japan), together with NRC (Canada), MOST and ASIAA (Taiwan), and KASI (Republic of Korea), in cooperation with the Republic of Chile. The Joint ALMA Observatory is operated by ESO, AUI/NRAO and NAOJ. The National Radio Astronomy Observatory is a facility of the National Science Foundation operated under cooperative agreement by Associated Universities, Inc. A.G. and T.W. acknowledge support from NSF AAG award 2009849. A.B. acknowledges support from NSF AAG award 2108140. R.I acknowledges support from NSF AAG award 2009624. O.N. was supported by the Director's Discretionary Fund at the Space Telescope Science Institute and the NASA Postdoctoral Program at NASA Goddard Space Flight Center, administered by Oak Ridge Associated Universities under contract with NASA. M.R. acknowledges support from ANID(CHILE) through Basal FB210003 and FONDECYT grant No1190684. ASH is supported in part by an STScI Postdoctoral Fellowship. This research made use of \texttt{astrodendro}, a Python package to compute dendrograms of Astronomical data \footnote{\url{http://www.dendrograms.org/}}; \texttt{SCIMES}, a Python package to find relevant structures into dendrograms of molecular gas emission using the spectral clustering approach \citep{2015MNRAS.454.2067C}; Astropy,\footnote{\url{http://www.astropy.org}} a community-developed core Python package for Astronomy \citep{astropy:2013, astropy:2018, astropy:2022}; and APLpy, an open-source plotting package for Python \citep{Robitaille_2012}. 
\end{acknowledgments}
\appendix

\section{Correlations between star-formation tracers and the residuals of the size-linewidth relation fits} \label{appendix:correlation}
\begin{figure*}[!htb] 
    \centering
    \includegraphics[width=\textwidth, trim={2.7in 0.45in 3.19in 1.045in},clip]{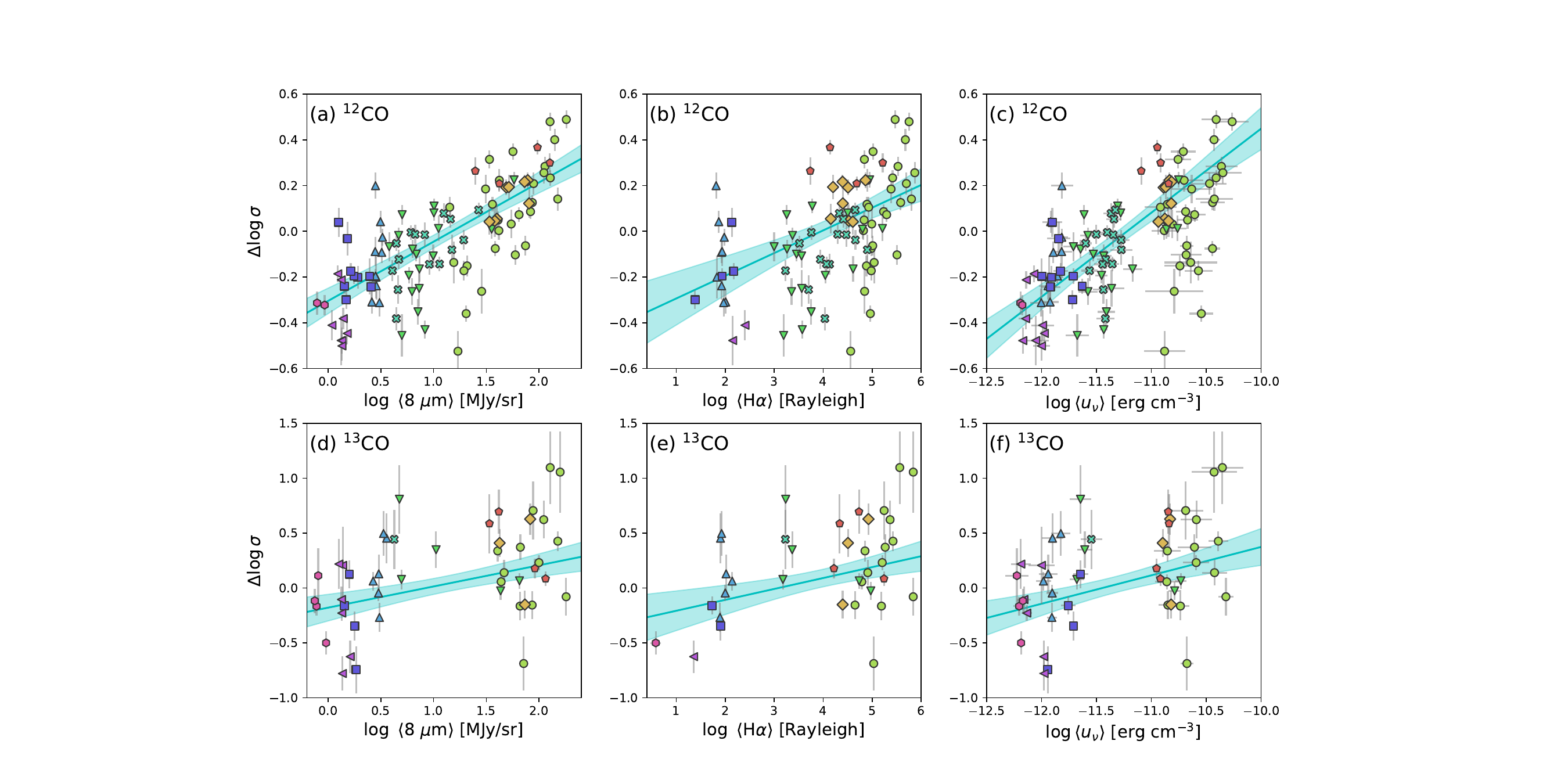}
    \caption{Star-formation tracers vs.\ $\Delta \log \sigma$ for (a-c) $^{12}$CO and (d-f) $^{13}$CO clumps, following the layout of Figure \ref{fig:nu0}. The fits are similar to the star-formation tracer vs.\ $\log \nu_0$ fits, though the $^{13}$CO data have weaker correlation, due to the poorly constrained $^{13}$CO size-linewidth fit.}
    \label{fig:dls_sft}
\end{figure*}

As demonstrated in Section \ref{sec:SLW_fits}, the difference between the fiducial and fitted slopes is $1.7\sigma$ and $2.3\sigma$ levels for $^{12}$CO and $^{13}$CO respectively. In the analysis conducted in Section \ref{sec:sft_correlations}, we correlated the star-formation tracers with the normalization of the fiducial relation. This normalization assumes that the fiducial size-linewidth relation is consistent with the data, and thus the correlation analysis in Section \ref{sec:sft_correlations} may be sensitive to the choice of slope. Motivated by the $\sim 2\sigma$ difference in slopes, we investigate the robustness of this correlation analysis to the assumed size-linewidth relation by referencing the linewidth to the fitted rather than fiducial relation. We define 
\begin{equation}
    \Delta \log \sigma \equiv \log \sigma_v - \log \sigma_{\text{fit}},
\end{equation}
the residual of the observed, deconvolved clump linewidths with respect to the radius power-law fits identified in Section \ref{sec:SLW_fits}. The uncertainty of $\Delta\log\sigma$ is the sum-in-quadrature of the uncertainties of the logarithm of the observed linewidth and of the $1\sigma$ confidence interval obtained from the ODR fitting. We plot $\Delta \log \sigma$ against star-formation tracers in Figure \ref{fig:dls_sft}, mirroring the conventions of Figure \ref{fig:nu0}. The fit parameters, scatter, and correlation coefficients are likewise presented in Table \ref{table:dls_params}.

The correlation coefficients of the star formation tracers vs.\ $\Delta\log\sigma$ for $^{12}$CO are very similar to those against $\log \nu_0$, but for $^{13}$CO, the correlation coefficients with $\Delta\log\sigma$ are much weaker. The $^{12}$CO fits with $\Delta\log\sigma$ have similar $\varepsilon$ to the corresponding $\log \nu_0$ fits, but have smaller $\varepsilon_{\text{norm}}$. The $^{13}$CO fits similarly have smaller $\varepsilon_{\text{norm}}$, though have much larger $\varepsilon$. Both $\varepsilon$ and $\rho$ measure goodness-of-fit unweighted by uncertainty, but the fits downweight data with larger uncertainty. Thus since the $^{13}$CO size-linewidth fit is more poorly constrained than the $^{12}$CO fit, it is not surprising that the $^{13}$CO star-formation tracer vs.\ $\Delta\log\sigma$ correlations appear poorer than the corresponding $\log\nu_0$ correlations with these metrics. $\varepsilon_{\text{norm}}$, on the other hand, more accurately reflects how meausrement uncertainties influence the size-linewidth fits. 

For both lines, all of the pairs of fits (star formation tracer vs.\ $\log\nu_0$ and vs.\ $\Delta\log\sigma$) have slopes consistent within $\sim 1\sigma$. While the significance of the fits are sensitive to the exact choice of slope of the size-linewidth relation, there does appear to be a correlation between star formation tracers and the normalization of both the fitted and fiducial size-linewidth relations.

\begin{deluxetable*}{r|r|r|r|r|r|r}[hbt!] \label{table:dls_params}
    \tablecaption{Parameters, scatter, and correlation coefficient of fits to the residual of the best fit relation. $m$ and $X_0$ are the parameters and uncertainties fit by ODR to the model $\Delta\log\sigma = m \left(X - X_0\right)$. The columns follow those of Table \ref{table:fiducial_relation}.}
    \tablehead{\colhead{} & \colhead{} & \colhead{$m$} & \colhead{X$_0$} & \colhead{$\varepsilon$} & \colhead{$\varepsilon_{\text{norm}}$} & \colhead{$\rho$}}
    \startdata
    \multirow{4}{*}{$^{12}$CO} & $\Delta\log\sigma$ & & & 0.24 & 8.10 & 0.34 \\
    & $\Delta \log \sigma$ vs.\ $\log\langle 8 \mu\text{m} \rangle$ & 0.26 $\pm$ 0.02 & 1.18 $\pm$ 0.05 & 0.16 & 3.24 & 0.74 \\
    & $\Delta \log \sigma$ vs.\ $\log\langle H\alpha \rangle$ & 0.10 $\pm$ 0.02 & 3.96 $\pm$ 0.19 & 0.19 & 4.08 & 0.51 \\
    & $\Delta \log \sigma$ vs.\ $\log\langle u_{\nu} \rangle$ & 0.37 $\pm$ 0.03 & -11.22 $\pm$ 0.04 & 0.19 & 3.99 & 0.65 \\
    \hline
    \multirow{4}{*}{$^{13}$CO} & $\Delta\log\sigma$ & & & 0.45 & 15.07 & 0.29 \\
    & $\Delta \log \sigma$ vs.\ $\log\langle 8 \mu \text{m} \rangle$ & 0.19 $\pm$ 0.04 & 0.93 $\pm$ 0.19 & 0.40 & 2.07 & 0.44 \\
    & $\Delta \log \sigma$ vs.\ $\log\langle H\alpha \rangle$ & 0.10 $\pm$ 0.03 & 3.09 $\pm$ 0.46 & 0.40 & 2.07 & 0.41 \\
    & $\Delta \log \sigma$ vs.\ $\log\langle u_{\nu} \rangle$ & 0.26 $\pm$ 0.06 & -11.45 $\pm$ 0.14 & 0.40 & 2.09 & 0.43
    \enddata
\end{deluxetable*}

\section{Correlations between clump-averaged star-formation tracers and surface densities} \label{appendix:tracer_correlations}
\begin{figure*}[!htb]
    \centering
    \includegraphics[width=\columnwidth, trim={0.65in, 0.77in, 0.94in, 1.33in},clip]{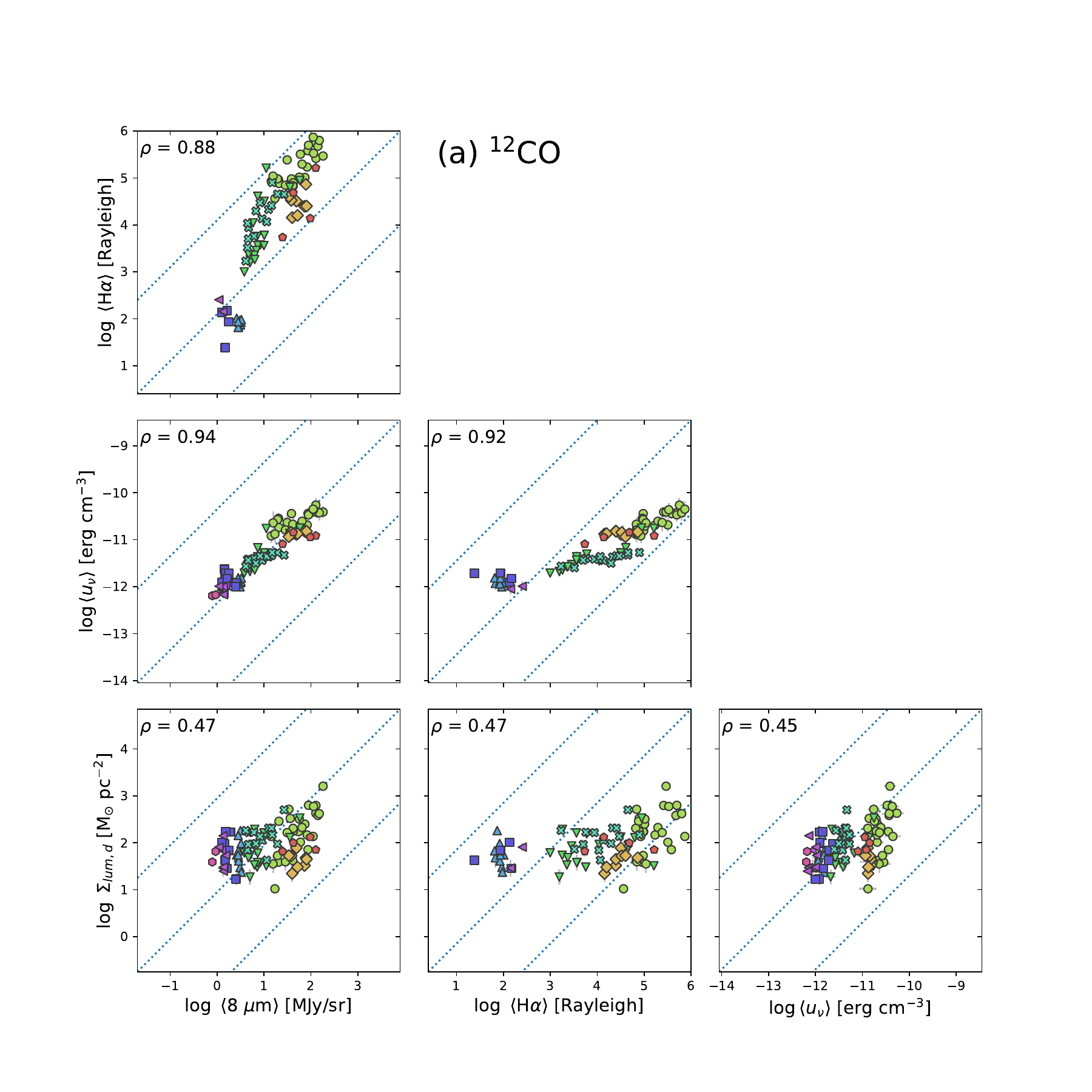} 
    \includegraphics[width=\columnwidth, trim={0.65in, 0.77in, 0.94in, 1.33in},clip]{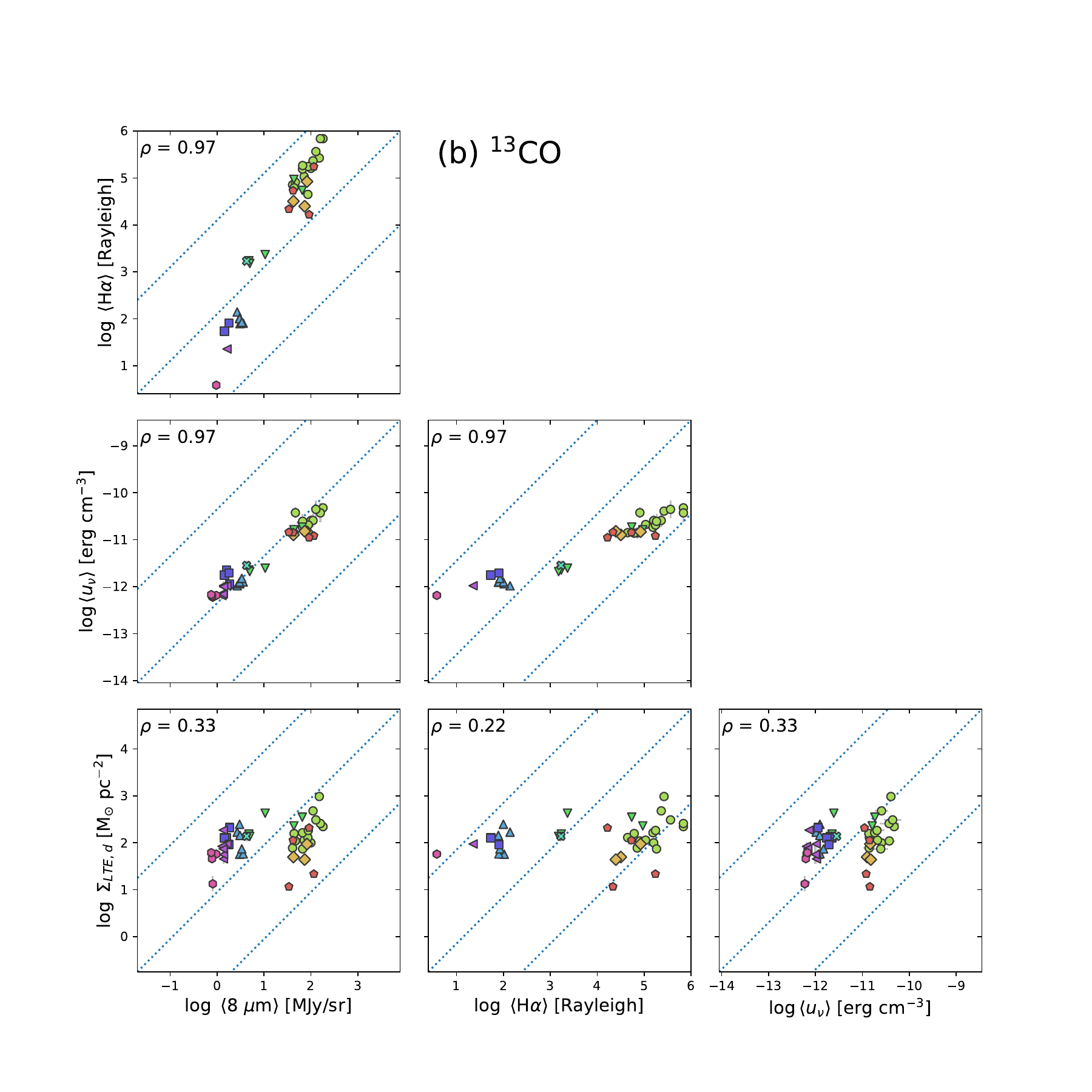} 
    \caption{ Correlations between the three star-formation tracers and CO-based surface density in (a) $^{12}$CO and (b) $^{13}$CO clumps. The correlation coefficients are inset in the top left of each panel. All axes have the same dynamic range, and the dotted blue lines mark relationships with unity slope.  }
    \label{fig:corner_correlations}
\end{figure*}

In Section \ref{sec:ir_optical_tracers}, we described the different processes traced by our ensemble of star-formation tracers. To show that these star-formation tracers are indicative of the strength of feedback mechanisms incident on clumps of molecular gas, we present the mutual correlations between the star-formation tracers, as well as clump surface densities, in Figure \ref{fig:corner_correlations}. All three tracers are strongly correlated with one another ($\rho \geq$ 0.88), but the relation between \avgH and the other two tracers is highly non-linear. The relationship between \avgm and \avgu is linear, which supports the supposition that $8\micron$ and $u_{\nu}$ trace the feedback from similar populations of stars, while H$\alpha$ is tracing evolutionary younger feedback sources. The correlations between the star-formation tracers and clump surface densities are much weaker, especially in $^{13}$CO.

\bibliographystyle{aasjournal}
\bibliography{lmc_clouds}{}

\begin{thebibliography}{}
\expandafter\ifx\csname natexlab\endcsname\relax\def\natexlab#1{#1}\fi
\providecommand{\url}[1]{\href{#1}{#1}}
\providecommand{\dodoi}[1]{doi:~\href{http://doi.org/#1}{\nolinkurl{#1}}}
\providecommand{\doeprint}[1]{\href{http://ascl.net/#1}{\nolinkurl{http://ascl.net/#1}}}
\providecommand{\doarXiv}[1]{\href{https://arxiv.org/abs/#1}{\nolinkurl{https://arxiv.org/abs/#1}}}

\bibitem[{Aniano {et~al.}(2011)Aniano, Draine, Gordon, \& Sandstrom}]{aniano_common-resolution_2011}
Aniano, G., Draine, B.~T., Gordon, K.~D., \& Sandstrom, K. 2011, \pasp, 123, 1218, \dodoi{10.1086/662219}

\bibitem[{{Astropy Collaboration} {et~al.}(2013){Astropy Collaboration}, {Robitaille}, {Tollerud}, {Greenfield}, {Droettboom}, {Bray}, {Aldcroft}, {Davis}, {Ginsburg}, {Price-Whelan}, {Kerzendorf}, {Conley}, {Crighton}, {Barbary}, {Muna}, {Ferguson}, {Grollier}, {Parikh}, {Nair}, {Unther}, {Deil}, {Woillez}, {Conseil}, {Kramer}, {Turner}, {Singer}, {Fox}, {Weaver}, {Zabalza}, {Edwards}, {Azalee Bostroem}, {Burke}, {Casey}, {Crawford}, {Dencheva}, {Ely}, {Jenness}, {Labrie}, {Lim}, {Pierfederici}, {Pontzen}, {Ptak}, {Refsdal}, {Servillat}, \& {Streicher}}]{astropy:2013}
{Astropy Collaboration}, {Robitaille}, T.~P., {Tollerud}, E.~J., {et~al.} 2013, \aap, 558, A33, \dodoi{10.1051/0004-6361/201322068}

\bibitem[{{Astropy Collaboration} {et~al.}(2018){Astropy Collaboration}, {Price-Whelan}, {Sip{\H{o}}cz}, {G{\"u}nther}, {Lim}, {Crawford}, {Conseil}, {Shupe}, {Craig}, {Dencheva}, {Ginsburg}, {Vand erPlas}, {Bradley}, {P{\'e}rez-Su{\'a}rez}, {de Val-Borro}, {Aldcroft}, {Cruz}, {Robitaille}, {Tollerud}, {Ardelean}, {Babej}, {Bach}, {Bachetti}, {Bakanov}, {Bamford}, {Barentsen}, {Barmby}, {Baumbach}, {Berry}, {Biscani}, {Boquien}, {Bostroem}, {Bouma}, {Brammer}, {Bray}, {Breytenbach}, {Buddelmeijer}, {Burke}, {Calderone}, {Cano Rodr{\'\i}guez}, {Cara}, {Cardoso}, {Cheedella}, {Copin}, {Corrales}, {Crichton}, {D'Avella}, {Deil}, {Depagne}, {Dietrich}, {Donath}, {Droettboom}, {Earl}, {Erben}, {Fabbro}, {Ferreira}, {Finethy}, {Fox}, {Garrison}, {Gibbons}, {Goldstein}, {Gommers}, {Greco}, {Greenfield}, {Groener}, {Grollier}, {Hagen}, {Hirst}, {Homeier}, {Horton}, {Hosseinzadeh}, {Hu}, {Hunkeler}, {Ivezi{\'c}}, {Jain}, {Jenness}, {Kanarek}, {Kendrew}, {Kern}, {Kerzendorf}, {Khvalko}, {King}, {Kirkby}, {Kulkarni},
  {Kumar}, {Lee}, {Lenz}, {Littlefair}, {Ma}, {Macleod}, {Mastropietro}, {McCully}, {Montagnac}, {Morris}, {Mueller}, {Mumford}, {Muna}, {Murphy}, {Nelson}, {Nguyen}, {Ninan}, {N{\"o}the}, {Ogaz}, {Oh}, {Parejko}, {Parley}, {Pascual}, {Patil}, {Patil}, {Plunkett}, {Prochaska}, {Rastogi}, {Reddy Janga}, {Sabater}, {Sakurikar}, {Seifert}, {Sherbert}, {Sherwood-Taylor}, {Shih}, {Sick}, {Silbiger}, {Singanamalla}, {Singer}, {Sladen}, {Sooley}, {Sornarajah}, {Streicher}, {Teuben}, {Thomas}, {Tremblay}, {Turner}, {Terr{\'o}n}, {van Kerkwijk}, {de la Vega}, {Watkins}, {Weaver}, {Whitmore}, {Woillez}, {Zabalza}, \& {Astropy Contributors}}]{astropy:2018}
{Astropy Collaboration}, {Price-Whelan}, A.~M., {Sip{\H{o}}cz}, B.~M., {et~al.} 2018, \aj, 156, 123, \dodoi{10.3847/1538-3881/aabc4f}

\bibitem[{{Astropy Collaboration} {et~al.}(2022){Astropy Collaboration}, {Price-Whelan}, {Lim}, {Earl}, {Starkman}, {Bradley}, {Shupe}, {Patil}, {Corrales}, {Brasseur}, {N{"o}the}, {Donath}, {Tollerud}, {Morris}, {Ginsburg}, {Vaher}, {Weaver}, {Tocknell}, {Jamieson}, {van Kerkwijk}, {Robitaille}, {Merry}, {Bachetti}, {G{"u}nther}, {Aldcroft}, {Alvarado-Montes}, {Archibald}, {B{'o}di}, {Bapat}, {Barentsen}, {Baz{'a}n}, {Biswas}, {Boquien}, {Burke}, {Cara}, {Cara}, {Conroy}, {Conseil}, {Craig}, {Cross}, {Cruz}, {D'Eugenio}, {Dencheva}, {Devillepoix}, {Dietrich}, {Eigenbrot}, {Erben}, {Ferreira}, {Foreman-Mackey}, {Fox}, {Freij}, {Garg}, {Geda}, {Glattly}, {Gondhalekar}, {Gordon}, {Grant}, {Greenfield}, {Groener}, {Guest}, {Gurovich}, {Handberg}, {Hart}, {Hatfield-Dodds}, {Homeier}, {Hosseinzadeh}, {Jenness}, {Jones}, {Joseph}, {Kalmbach}, {Karamehmetoglu}, {Ka{l}uszy{'n}ski}, {Kelley}, {Kern}, {Kerzendorf}, {Koch}, {Kulumani}, {Lee}, {Ly}, {Ma}, {MacBride}, {Maljaars}, {Muna}, {Murphy}, {Norman}, {O'Steen},
  {Oman}, {Pacifici}, {Pascual}, {Pascual-Granado}, {Patil}, {Perren}, {Pickering}, {Rastogi}, {Roulston}, {Ryan}, {Rykoff}, {Sabater}, {Sakurikar}, {Salgado}, {Sanghi}, {Saunders}, {Savchenko}, {Schwardt}, {Seifert-Eckert}, {Shih}, {Jain}, {Shukla}, {Sick}, {Simpson}, {Singanamalla}, {Singer}, {Singhal}, {Sinha}, {Sip{H{o}}cz}, {Spitler}, {Stansby}, {Streicher}, {{{S}}umak}, {Swinbank}, {Taranu}, {Tewary}, {Tremblay}, {Val-Borro}, {Van Kooten}, {Vasovi{'c}}, {Verma}, {de Miranda Cardoso}, {Williams}, {Wilson}, {Winkel}, {Wood-Vasey}, {Xue}, {Yoachim}, {Zhang}, {Zonca}, \& {Astropy Project Contributors}}]{astropy:2022}
{Astropy Collaboration}, {Price-Whelan}, A.~M., {Lim}, P.~L., {et~al.} 2022, apj, 935, 167, \dodoi{10.3847/1538-4357/ac7c74}

\bibitem[{Ballesteros-Paredes {et~al.}(2011)Ballesteros-Paredes, Hartmann, Vázquez-Semadeni, Heitsch, \& Zamora-Avilés}]{ballesteros-paredes_gravity_2011}
Ballesteros-Paredes, J., Hartmann, L.~W., Vázquez-Semadeni, E., Heitsch, F., \& Zamora-Avilés, M.~A. 2011, \mnras, 411, 65, \dodoi{10.1111/j.1365-2966.2010.17657.x}

\bibitem[{{Ballesteros-Paredes} \& {Mac Low}(2002)}]{ballesteros_paredes_physical_2002}
{Ballesteros-Paredes}, J., \& {Mac Low}, M.-M. 2002, \apj, 570, 734, \dodoi{10.1086/339624}

\bibitem[{Bally(2016)}]{bally_protostellar_2016}
Bally, J. 2016, \araa, 54, 491, \dodoi{10.1146/annurev-astro-081915-023341}

\bibitem[{{Barnes} {et~al.}(2021){Barnes}, {Glover}, {Kreckel}, {Ostriker}, {Bigiel}, {Belfiore}, {Be{\v{s}}li{\'c}}, {Blanc}, {Chevance}, {Dale}, {Egorov}, {Eibensteiner}, {Emsellem}, {Grasha}, {Groves}, {Klessen}, {Kruijssen}, {Leroy}, {Longmore}, {Lopez}, {McElroy}, {Meidt}, {Murphy}, {Rosolowsky}, {Saito}, {Santoro}, {Schinnerer}, {Schruba}, {Sun}, {Watkins}, \& {Williams}}]{barnes_comparing_2021}
{Barnes}, A.~T., {Glover}, S.~C.~O., {Kreckel}, K., {et~al.} 2021, \mnras, 508, 5362, \dodoi{10.1093/mnras/stab2958}

\bibitem[{Bendo {et~al.}(2008)Bendo, Draine, Engelbracht, Helou, Thornley, Bot, Buckalew, Calzetti, Dale, Hollenbach, Li, \& Moustakas}]{bendo_relations_2008}
Bendo, G.~J., Draine, B.~T., Engelbracht, C.~W., {et~al.} 2008, \mnras, 389, 629, \dodoi{10.1111/j.1365-2966.2008.13567.x}

\bibitem[{Bolatto {et~al.}(2008)Bolatto, Leroy, Rosolowsky, Walter, \& Blitz}]{bolatto_resolved_2008}
Bolatto, A.~D., Leroy, A.~K., Rosolowsky, E., Walter, F., \& Blitz, L. 2008, \apj, 686, 948, \dodoi{10.1086/591513}

\bibitem[{Calzetti {et~al.}(2007)Calzetti, Kennicutt, Engelbracht, Leitherer, Draine, Kewley, Moustakas, Sosey, Dale, Gordon, Helou, Hollenbach, Armus, Bendo, Bot, Buckalew, Jarrett, Li, Meyer, Murphy, Prescott, Regan, Rieke, Roussel, Sheth, Smith, Thornley, \& Walter}]{calzetti_calibration_2007}
Calzetti, D., Kennicutt, R.~C., Engelbracht, C.~W., {et~al.} 2007, \apj, 666, 870, \dodoi{10.1086/520082}

\bibitem[{Chevance {et~al.}(2023)Chevance, Krumholz, McLeod, Ostriker, Rosolowsky, \& Sternberg}]{chevance_life_2023}
Chevance, M., Krumholz, M.~R., McLeod, A.~F., {et~al.} 2023, 534, 1, \dodoi{10.48550/arXiv.2203.09570}

\bibitem[{Chiang {et~al.}(2018)Chiang, Sandstrom, Chastenet, Johnson, Leroy, \& Utomo}]{chiang_spatially_2018}
Chiang, I.-D., Sandstrom, K.~M., Chastenet, J., {et~al.} 2018, \apj, 865, 117, \dodoi{10.3847/1538-4357/aadc5f}

\bibitem[{Choudhury {et~al.}(2021)Choudhury, de~Grijs, Bekki, Cioni, Ivanov, van Loon, Miller, Niederhofer, Oliveira, Ripepi, Sun, \& Subramanian}]{choudhury_vmc_2021}
Choudhury, S., de~Grijs, R., Bekki, K., {et~al.} 2021, \mnras, 507, 4752, \dodoi{10.1093/mnras/stab2446}

\bibitem[{Cioni(2009)}]{cioni_metallicity_2009}
Cioni, M. R.~L. 2009, Astronomy and Astrophysics, 506, 1137, \dodoi{10.1051/0004-6361/200912138}

\bibitem[{Colombo {et~al.}(2015)Colombo, Rosolowsky, Ginsburg, Duarte-Cabral, \& Hughes}]{colombo_graph-based_2015}
Colombo, D., Rosolowsky, E., Ginsburg, A., Duarte-Cabral, A., \& Hughes, A. 2015, \mnras, 454, 2067, \dodoi{10.1093/mnras/stv2063}

\bibitem[{{Colombo} {et~al.}(2015){Colombo}, {Rosolowsky}, {Ginsburg}, {Duarte-Cabral}, \& {Hughes}}]{2015MNRAS.454.2067C}
{Colombo}, D., {Rosolowsky}, E., {Ginsburg}, A., {Duarte-Cabral}, A., \& {Hughes}, A. 2015, \mnras, 454, 2067, \dodoi{10.1093/mnras/stv2063}

\bibitem[{Cortzen {et~al.}(2019)Cortzen, Garrett, Magdis, Rigopoulou, Valentino, Pereira-Santaella, Combes, Alonso-Herrero, Toft, Daddi, Elbaz, Gómez-Guijarro, Stockmann, Huang, \& Kramer}]{cortzen_pahs_2019}
Cortzen, I., Garrett, J., Magdis, G., {et~al.} 2019, \mnras, 482, 1618, \dodoi{10.1093/mnras/sty2777}

\bibitem[{Draine \& Li(2007)}]{draine_infrared_2007}
Draine, B.~T., \& Li, A. 2007, \apj, 657, 810, \dodoi{10.1086/511055}

\bibitem[{Emerick {et~al.}(2018)Emerick, Bryan, \& Mac~Low}]{emerick_stellar_2018}
Emerick, A., Bryan, G.~L., \& Mac~Low, M.-M. 2018, \apj, 865, L22, \dodoi{10.3847/2041-8213/aae315}

\bibitem[{Field {et~al.}(2011)Field, Blackman, \& Keto}]{field_does_2011}
Field, G.~B., Blackman, E.~G., \& Keto, E.~R. 2011, \mnras, 416, 710, \dodoi{10.1111/j.1365-2966.2011.19091.x}

\bibitem[{Finn {et~al.}(2021)Finn, Indebetouw, Johnson, Costa, Chen, Kawamura, Onishi, Ott, Tokuda, Wong, \& Zahorecz}]{finn_physical_2021}
Finn, M.~K., Indebetouw, R., Johnson, K.~E., {et~al.} 2021, ApJ, 917, 106, \dodoi{10.3847/1538-4357/ac090c}

\bibitem[{Fukui \& Kawamura(2010)}]{fukui_molecular_2010}
Fukui, Y., \& Kawamura, A. 2010, \araa, 48, 547, \dodoi{10.1146/annurev-astro-081309-130854}

\bibitem[{{Fukui} \& {Kawamura}(2010)}]{2010ARA&A..48..547F}
{Fukui}, Y., \& {Kawamura}, A. 2010, \araa, 48, 547, \dodoi{10.1146/annurev-astro-081309-130854}

\bibitem[{Fukui {et~al.}(2015)Fukui, Harada, Tokuda, Morioka, Onishi, Torii, Ohama, Hattori, Nayak, Meixner, Sewiło, Indebetouw, Kawamura, Saigo, Yamamoto, Tachihara, Minamidani, Inoue, Madden, Galametz, Lebouteiller, Mizuno, \& Chen}]{fukui_high-mass_2015}
Fukui, Y., Harada, R., Tokuda, K., {et~al.} 2015, \apj, 807, L4, \dodoi{10.1088/2041-8205/807/1/L4}

\bibitem[{{Gonz{\'a}lez Lobos} \& {Stutz}(2019)}]{2019MNRAS.489.4771G}
{Gonz{\'a}lez Lobos}, V., \& {Stutz}, A.~M. 2019, \mnras, 489, 4771, \dodoi{10.1093/mnras/stz2512}

\bibitem[{{Gordon}(2018)}]{gordonthesis2018}
{Gordon}, A. J.~R. 2018, {A Narrowband Emission-Line Survey of the Large Magellanic Cloud}

\bibitem[{Gordon {et~al.}(2014)Gordon, Roman-Duval, Bot, Meixner, Babler, Bernard, Bolatto, Boyer, Clayton, Engelbracht, Fukui, Galametz, Galliano, Hony, Hughes, Indebetouw, Israel, Jameson, Kawamura, Lebouteiller, Li, Madden, Matsuura, Misselt, Montiel, Okumura, Onishi, Panuzzo, Paradis, Rubio, Sandstrom, Sauvage, Seale, Sewiło, Tchernyshyov, \& Skibba}]{gordon_dust_2014}
Gordon, K.~D., Roman-Duval, J., Bot, C., {et~al.} 2014, \apj, 797, 85, \dodoi{10.1088/0004-637X/797/2/85}

\bibitem[{Grady {et~al.}(2021)Grady, Belokurov, \& Evans}]{grady_magellanic_2021}
Grady, J., Belokurov, V., \& Evans, N.~W. 2021, \apj, 909, 150, \dodoi{10.3847/1538-4357/abd4e4}

\bibitem[{Gritschneder {et~al.}(2009)Gritschneder, Naab, Walch, Burkert, \& Heitsch}]{gritschneder_driving_2009}
Gritschneder, M., Naab, T., Walch, S., Burkert, A., \& Heitsch, F. 2009, ApJ, 694, L26, \dodoi{10.1088/0004-637X/694/1/L26}

\bibitem[{Grudić {et~al.}(2022)Grudić, Guszejnov, Offner, Rosen, Raju, Faucher-Giguère, \& Hopkins}]{grudic_dynamics_2022}
Grudić, M.~Y., Guszejnov, D., Offner, S. S.~R., {et~al.} 2022, \mnras, 512, 216, \dodoi{10.1093/mnras/stac526}

\bibitem[{Hacar {et~al.}(2016)Hacar, Alves, Burkert, \& Goldsmith}]{hacar_opacity_2016}
Hacar, A., Alves, J., Burkert, A., \& Goldsmith, P. 2016, Astronomy and Astrophysics, 591, A104, \dodoi{10.1051/0004-6361/201527319}

\bibitem[{Haid {et~al.}(2018)Haid, Walch, Seifried, Wünsch, Dinnbier, \& Naab}]{haid_relative_2018}
Haid, S., Walch, S., Seifried, D., {et~al.} 2018, \mnras, 478, 4799, \dodoi{10.1093/mnras/sty1315}

\bibitem[{Haschke {et~al.}(2011)Haschke, Grebel, \& Duffau}]{haschke_new_2011}
Haschke, R., Grebel, E.~K., \& Duffau, S. 2011, The Astronomical Journal, 141, 158, \dodoi{10.1088/0004-6256/141/5/158}

\bibitem[{Hennebelle \& Falgarone(2012)}]{hennebelle_turbulent_2012}
Hennebelle, P., \& Falgarone, E. 2012, Astronomy and Astrophysics Review, 20, 55, \dodoi{10.1007/s00159-012-0055-y}

\bibitem[{Heyer \& Dame(2015)}]{heyer_molecular_2015}
Heyer, M., \& Dame, T.~M. 2015, \araa, 53, 583, \dodoi{10.1146/annurev-astro-082214-122324}

\bibitem[{Heyer {et~al.}(2009)Heyer, Krawczyk, Duval, \& Jackson}]{heyer_re-examining_2009}
Heyer, M., Krawczyk, C., Duval, J., \& Jackson, J.~M. 2009, \apj, 699, 1092, \dodoi{10.1088/0004-637X/699/2/1092}

\bibitem[{Hopkins {et~al.}(2011)Hopkins, Quataert, \& Murray}]{hopkins_self-regulated_2011}
Hopkins, P.~F., Quataert, E., \& Murray, N. 2011, \mnras, 417, 950, \dodoi{10.1111/j.1365-2966.2011.19306.x}

\bibitem[{Hughes {et~al.}(2010)Hughes, Wong, Ott, Muller, Pineda, Mizuno, Bernard, Paradis, Maddison, Reach, Staveley-Smith, Kawamura, Meixner, Kim, Onishi, Mizuno, \& Fukui}]{hughes_physical_2010}
Hughes, A., Wong, T., Ott, J., {et~al.} 2010, \mnras, 406, 2065, \dodoi{10.1111/j.1365-2966.2010.16829.x}

\bibitem[{Indebetouw {et~al.}(2013)Indebetouw, Brogan, Chen, Leroy, Johnson, Muller, Madden, Cormier, Galliano, Hughes, Hunter, Kawamura, Kepley, Lebouteiller, Meixner, Oliveira, Onishi, \& Vasyunina}]{indebetouw_alma_2013}
Indebetouw, R., Brogan, C., Chen, C. H.~R., {et~al.} 2013, \apj, 774, 73, \dodoi{10.1088/0004-637X/774/1/73}

\bibitem[{{Kavak} {et~al.}(2022){Kavak}, {Bally}, {Goicoechea}, {Pabst}, {van der Tak}, \& {Tielens}}]{kavak_dents_2022}
{Kavak}, {\"U}., {Bally}, J., {Goicoechea}, J.~R., {et~al.} 2022, \aap, 663, A117, \dodoi{10.1051/0004-6361/202243332}

\bibitem[{Kennicutt \& Evans(2012)}]{kennicutt_star_2012}
Kennicutt, R.~C., \& Evans, N.~J. 2012, \araa, 50, 531, \dodoi{10.1146/annurev-astro-081811-125610}

\bibitem[{Klessen \& Hennebelle(2010)}]{klessen_accretion-driven_2010}
Klessen, R.~S., \& Hennebelle, P. 2010, A\&A, 520, A17, \dodoi{10.1051/0004-6361/200913780}

\bibitem[{Koo {et~al.}(2020)Koo, Kim, Park, \& Ostriker}]{koo_radiative_2020}
Koo, B.-C., Kim, C.-G., Park, S., \& Ostriker, E.~C. 2020, \apj, 905, 35, \dodoi{10.3847/1538-4357/abc1e7}

\bibitem[{Larson(1981)}]{larson_turbulence_1981}
Larson, R.~B. 1981, \mnras, 194, 809, \dodoi{10.1093/mnras/194.4.809}

\bibitem[{Lee {et~al.}(2016)Lee, Miville-Deschênes, \& Murray}]{lee_observational_2016}
Lee, E.~J., Miville-Deschênes, M.-A., \& Murray, N.~W. 2016, \apj, 833, 229, \dodoi{10.3847/1538-4357/833/2/229}

\bibitem[{Leroy {et~al.}(2016)Leroy, Hughes, Schruba, Rosolowsky, Blanc, Bolatto, Colombo, Escala, Kramer, Kruijssen, Meidt, Pety, Querejeta, Sandstrom, Schinnerer, Sliwa, \& Usero}]{leroy_portrait_2016}
Leroy, A.~K., Hughes, A., Schruba, A., {et~al.} 2016, \apj, 831, 16, \dodoi{10.3847/0004-637X/831/1/16}

\bibitem[{Leroy {et~al.}(2023)Leroy, Bolatto, Sandstrom, Rosolowsky, Barnes, Bigiel, Boquien, den Brok, Cao, Chastenet, Chevance, Chiang, Chown, Colombo, Ellison, Emsellem, Grasha, Henshaw, Hughes, Klessen, Koch, Kim, Kreckel, Kruijssen, Larson, Lee, Levy, Lin, Liu, Meidt, Pety, Querejeta, Rubio, Saito, Salim, Schinnerer, Sormani, Sun, Thilker, Usero, Vogel, Watkins, Whitcomb, Williams, \& Wilson}]{leroy_phangsjwst_2023}
Leroy, A.~K., Bolatto, A.~D., Sandstrom, K., {et~al.} 2023, ApJL, 944, L10, \dodoi{10.3847/2041-8213/acab01}

\bibitem[{Li \& Draine(2002)}]{li_infrared_2002}
Li, A., \& Draine, B.~T. 2002, \apj, 572, 232, \dodoi{10.1086/340285}

\bibitem[{Li {et~al.}(2023)Li, Huang, Liu, Beers, \& Zhang}]{li_photometric_2023}
Li, X.-Y., Huang, Y., Liu, G.-C., Beers, T.~C., \& Zhang, H.-W. 2023, \apj, 944, 88, \dodoi{10.3847/1538-4357/acadd5}

\bibitem[{Lopez {et~al.}(2014)Lopez, Krumholz, Bolatto, Prochaska, Ramirez-Ruiz, \& Castro}]{lopez_role_2014}
Lopez, L.~A., Krumholz, M.~R., Bolatto, A.~D., {et~al.} 2014, \apj, 795, 121, \dodoi{10.1088/0004-637X/795/2/121}

\bibitem[{Luck \& Lambert(2011)}]{luck_distribution_2011}
Luck, R.~E., \& Lambert, D.~L. 2011, The Astronomical Journal, 142, 136, \dodoi{10.1088/0004-6256/142/4/136}

\bibitem[{McKee \& Ostriker(2007)}]{mckee_theory_2007}
McKee, C.~F., \& Ostriker, E.~C. 2007, \araa, 45, 565, \dodoi{10.1146/annurev.astro.45.051806.110602}

\bibitem[{Meixner {et~al.}(2006)Meixner, Gordon, Indebetouw, Hora, Whitney, Blum, Reach, Bernard, Meade, Babler, Engelbracht, For, Misselt, Vijh, Leitherer, Cohen, Churchwell, Boulanger, Frogel, Fukui, Gallagher, Gorjian, Harris, Kelly, Kawamura, Kim, Latter, Madden, Markwick-Kemper, Mizuno, Mizuno, Mould, Nota, Oey, Olsen, Onishi, Paladini, Panagia, Perez-Gonzalez, Shibai, Sato, Smith, Staveley-Smith, Tielens, Ueta, van Dyk, Volk, Werner, \& Zaritsky}]{meixner_spitzer_2006}
Meixner, M., Gordon, K.~D., Indebetouw, R., {et~al.} 2006, The Astronomical Journal, 132, 2268, \dodoi{10.1086/508185}

\bibitem[{{Meixner} {et~al.}(2006){Meixner}, {Gordon}, {Indebetouw}, {Hora}, {Whitney}, {Blum}, {Reach}, {Bernard}, {Meade}, {Babler}, {Engelbracht}, {For}, {Misselt}, {Vijh}, {Leitherer}, {Cohen}, {Churchwell}, {Boulanger}, {Frogel}, {Fukui}, {Gallagher}, {Gorjian}, {Harris}, {Kelly}, {Kawamura}, {Kim}, {Latter}, {Madden}, {Markwick-Kemper}, {Mizuno}, {Mizuno}, {Mould}, {Nota}, {Oey}, {Olsen}, {Onishi}, {Paladini}, {Panagia}, {Perez-Gonzalez}, {Shibai}, {Sato}, {Smith}, {Staveley-Smith}, {Tielens}, {Ueta}, {van Dyk}, {Volk}, {Werner}, \& {Zaritsky}}]{2006AJ....132.2268M}
{Meixner}, M., {Gordon}, K.~D., {Indebetouw}, R., {et~al.} 2006, \aj, 132, 2268, \dodoi{10.1086/508185}

\bibitem[{Mestel \& Spitzer(1956)}]{mestel_star_1956}
Mestel, L., \& Spitzer, Jr., L. 1956, \mnras, 116, 503, \dodoi{10.1093/mnras/116.5.503}

\bibitem[{{Murray}(2011)}]{murray_star_2011}
{Murray}, N. 2011, \apj, 729, 133, \dodoi{10.1088/0004-637X/729/2/133}

\bibitem[{{Naslim} {et~al.}(2018){Naslim}, {Tokuda}, {Onishi}, {Kemper}, {Wong}, {Morata}, {Takada}, {Harada}, {Kawamura}, {Saigo}, {Indebetouw}, {Madden}, {Hony}, \& {Meixner}}]{2018ApJ...853..175N}
{Naslim}, N., {Tokuda}, K., {Onishi}, T., {et~al.} 2018, \apj, 853, 175, \dodoi{10.3847/1538-4357/aaa5b0}

\bibitem[{Naslim {et~al.}(2018)Naslim, Tokuda, Onishi, Kemper, Wong, Morata, Takada, Harada, Kawamura, Saigo, Indebetouw, Madden, Hony, \& Meixner}]{naslim_alma_2018}
Naslim, N., Tokuda, K., Onishi, T., {et~al.} 2018, \apj, 853, 175, \dodoi{10.3847/1538-4357/aaa5b0}

\bibitem[{Nayak {et~al.}(2018)Nayak, Meixner, Fukui, Tachihara, Onishi, Saigo, Tokuda, \& Harada}]{nayak_molecular_2018}
Nayak, O., Meixner, M., Fukui, Y., {et~al.} 2018, \apj, 854, 154, \dodoi{10.3847/1538-4357/aaab5f}

\bibitem[{Ochsendorf {et~al.}(2017)Ochsendorf, Meixner, Roman-Duval, Rahman, \& Evans}]{ochsendorf_what_2017}
Ochsendorf, B.~B., Meixner, M., Roman-Duval, J., Rahman, M., \& Evans, II, N.~J. 2017, \apj, 841, 109, \dodoi{10.3847/1538-4357/aa704a}

\bibitem[{{Pabst} {et~al.}(2019){Pabst}, {Higgins}, {Goicoechea}, {Teyssier}, {Berne}, {Chambers}, {Wolfire}, {Suri}, {Guesten}, {Stutzki}, {Graf}, {Risacher}, \& {Tielens}}]{pabst_disruption_2019}
{Pabst}, C., {Higgins}, R., {Goicoechea}, J.~R., {et~al.} 2019, \nat, 565, 618, \dodoi{10.1038/s41586-018-0844-1}

\bibitem[{Pedicelli {et~al.}(2009)Pedicelli, Bono, Lemasle, François, Groenewegen, Lub, Pel, Laney, Piersimoni, Romaniello, Buonanno, Caputo, Cassisi, Castelli, Leurini, Pietrinferni, Primas, \& Pritchard}]{pedicelli_metallicity_2009}
Pedicelli, S., Bono, G., Lemasle, B., {et~al.} 2009, Astronomy and Astrophysics, 504, 81, \dodoi{10.1051/0004-6361/200912504}

\bibitem[{Pietrzyński {et~al.}(2019)Pietrzyński, Graczyk, Gallenne, Gieren, Thompson, Pilecki, Karczmarek, Górski, Suchomska, Taormina, Zgirski, Wielgórski, Kołaczkowski, Konorski, Villanova, Nardetto, Kervella, Bresolin, Kudritzki, Storm, Smolec, \& Narloch}]{pietrzynski_distance_2019}
Pietrzyński, G., Graczyk, D., Gallenne, A., {et~al.} 2019, Nature, 567, 200, \dodoi{10.1038/s41586-019-0999-4}

\bibitem[{{Robitaille} \& {Bressert}(2012)}]{Robitaille_2012}
{Robitaille}, T., \& {Bressert}, E. 2012, {APLpy: Astronomical Plotting Library in Python}, Astrophysics Source Code Library, record ascl:1208.017.
\newblock \doeprint{1208.017}

\bibitem[{Rosolowsky \& Leroy(2006)}]{rosolowsky_bias-free_2006}
Rosolowsky, E., \& Leroy, A. 2006, \pasp, 118, 590, \dodoi{10.1086/502982}

\bibitem[{{Rosolowsky} {et~al.}(2008){Rosolowsky}, {Pineda}, {Kauffmann}, \& {Goodman}}]{2008ApJ...679.1338R}
{Rosolowsky}, E.~W., {Pineda}, J.~E., {Kauffmann}, J., \& {Goodman}, A.~A. 2008, \apj, 679, 1338, \dodoi{10.1086/587685}

\bibitem[{{SAGE Team}(2020)}]{https://doi.org/10.26131/irsa404}
{SAGE Team}. 2020, Surveying the Agents of a Galaxy's Evolution,  IPAC, \dodoi{10.26131/IRSA404}.
\newblock \url{https://catcopy.ipac.caltech.edu/dois/doi.php?id=10.26131/IRSA404}

\bibitem[{Saigo {et~al.}(2017)Saigo, Onishi, Nayak, Meixner, Tokuda, Harada, Morioka, Sewiło, Indebetouw, Torii, Kawamura, Ohama, Hattori, Yamamoto, Tachihara, Minamidani, Inoue, Madden, Galametz, Lebouteiller, Chen, Mizuno, \& Fukui}]{saigo_kinematic_2017}
Saigo, K., Onishi, T., Nayak, O., {et~al.} 2017, \apj, 835, 108, \dodoi{10.3847/1538-4357/835/1/108}

\bibitem[{{Schneider} \& {Brooks}(2004)}]{schneider_bias_2004}
{Schneider}, N., \& {Brooks}, K. 2004, \pasa, 21, 290, \dodoi{10.1071/AS04012}

\bibitem[{{Simon} {et~al.}(2001){Simon}, {Jackson}, {Clemens}, {Bania}, \& {Heyer}}]{simon_structure_2001}
{Simon}, R., {Jackson}, J.~M., {Clemens}, D.~P., {Bania}, T.~M., \& {Heyer}, M.~H. 2001, \apj, 551, 747, \dodoi{10.1086/320230}

\bibitem[{Smith \& {MCELS Team}(1998)}]{smith_umctio_1998}
Smith, R.~C., \& {MCELS Team}. 1998, \pasa, 15, 163, \dodoi{10.1071/AS98163}

\bibitem[{Solomon {et~al.}(1987)Solomon, Rivolo, Barrett, \& Yahil}]{solomon_mass_1987}
Solomon, P.~M., Rivolo, A.~R., Barrett, J., \& Yahil, A. 1987, \apj, 319, 730, \dodoi{10.1086/165493}

\bibitem[{Sorai {et~al.}(2001)Sorai, Hasegawa, Booth, Rubio, Morino, Bronfman, Handa, Hayashi, Nyman, Oka, Sakamoto, Seta, \& Usuda}]{sorai_co_2001}
Sorai, K., Hasegawa, T., Booth, R.~S., {et~al.} 2001, \apj, 551, 794, \dodoi{10.1086/320212}

\bibitem[{Sun {et~al.}(2018)Sun, Leroy, Schruba, Rosolowsky, Hughes, Kruijssen, Meidt, Schinnerer, Blanc, Bigiel, Bolatto, Chevance, Groves, Herrera, Hygate, Pety, Querejeta, Usero, \& Utomo}]{sun_cloud-scale_2018}
Sun, J., Leroy, A.~K., Schruba, A., {et~al.} 2018, \apj, 860, 172, \dodoi{10.3847/1538-4357/aac326}

\bibitem[{Sun {et~al.}(2020)Sun, Leroy, Schinnerer, Hughes, Rosolowsky, Querejeta, Schruba, Liu, Saito, Herrera, Faesi, Usero, Pety, Kruijssen, Ostriker, Bigiel, Blanc, Bolatto, Boquien, Chevance, Dale, Deger, Emsellem, Glover, Grasha, Groves, Henshaw, Jimenez-Donaire, Kim, Klessen, Kreckel, Lee, Meidt, Sandstrom, Sardone, Utomo, \& Williams}]{sun_molecular_2020}
Sun, J., Leroy, A.~K., Schinnerer, E., {et~al.} 2020, \apj, 901, L8, \dodoi{10.3847/2041-8213/abb3be}

\bibitem[{{Tiwari} {et~al.}(2021){Tiwari}, {Karim}, {Pound}, {Wolfire}, {Jacob}, {Buchbender}, {G{\"u}sten}, {Guevara}, {Higgins}, {Kabanovic}, {Pabst}, {Ricken}, {Schneider}, {Simon}, {Stutzki}, \& {Tielens}}]{tiwari_sofia_2021}
{Tiwari}, M., {Karim}, R., {Pound}, M.~W., {et~al.} 2021, \apj, 914, 117, \dodoi{10.3847/1538-4357/abf6ce}

\bibitem[{Utomo {et~al.}(2019)Utomo, Chiang, Leroy, Sandstrom, \& Chastenet}]{utomo_resolved_2019}
Utomo, D., Chiang, I.-D., Leroy, A.~K., Sandstrom, K.~M., \& Chastenet, J. 2019, \apj, 874, 141, \dodoi{10.3847/1538-4357/ab05d3}

\bibitem[{Utomo {et~al.}(2018)Utomo, Sun, Leroy, Kruijssen, Schinnerer, Schruba, Bigiel, Blanc, Chevance, Emsellem, Herrera, Hygate, Kreckel, Ostriker, Pety, Querejeta, Rosolowsky, Sandstrom, \& Usero}]{utomo_star_2018}
Utomo, D., Sun, J., Leroy, A.~K., {et~al.} 2018, \apj, 861, L18, \dodoi{10.3847/2041-8213/aacf8f}

\bibitem[{van~der Tak {et~al.}(2007)van~der Tak, Black, Schöier, Jansen, \& van Dishoeck}]{van_der_tak_computer_2007}
van~der Tak, F. F.~S., Black, J.~H., Schöier, F.~L., Jansen, D.~J., \& van Dishoeck, E.~F. 2007, Astronomy and Astrophysics, 468, 627, \dodoi{10.1051/0004-6361:20066820}

\bibitem[{Virtanen {et~al.}(2020)Virtanen, Gommers, Oliphant, Haberland, Reddy, Cournapeau, Burovski, Peterson, Weckesser, Bright, {van der Walt}, Brett, Wilson, Millman, Mayorov, Nelson, Jones, Kern, Larson, Carey, Polat, Feng, Moore, {VanderPlas}, Laxalde, Perktold, Cimrman, Henriksen, Quintero, Harris, Archibald, Ribeiro, Pedregosa, {van Mulbregt}, \& {SciPy 1.0 Contributors}}]{2020SciPy-NMeth}
Virtanen, P., Gommers, R., Oliphant, T.~E., {et~al.} 2020, Nature Methods, 17, 261, \dodoi{10.1038/s41592-019-0686-2}

\bibitem[{Walch {et~al.}(2015)Walch, Girichidis, Naab, Gatto, Glover, Wünsch, Klessen, Clark, Peters, Derigs, \& Baczynski}]{walch_silcc_2015}
Walch, S., Girichidis, P., Naab, T., {et~al.} 2015, \mnras, 454, 238, \dodoi{10.1093/mnras/stv1975}

\bibitem[{{Westerlund}(1990)}]{1990A&ARv...2...29W}
{Westerlund}, B.~E. 1990, \aapr, 2, 29, \dodoi{10.1007/BF00873541}

\bibitem[{Westerlund(1997)}]{westerlund_magellanic_1997}
Westerlund, B.~E. 1997, The {Magellanic} {Clouds}.
\newblock \url{https://ui.adsabs.harvard.edu/abs/1997macl.book.....W}

\bibitem[{Whitcomb {et~al.}(2023)Whitcomb, Sandstrom, Leroy, \& Smith}]{whitcomb_star_2023}
Whitcomb, C.~M., Sandstrom, K., Leroy, A., \& Smith, J. D.~T. 2023, \apj, 948, 88, \dodoi{10.3847/1538-4357/acc316}

\bibitem[{Wise {et~al.}(2012)Wise, Abel, Turk, Norman, \& Smith}]{wise_birth_2012}
Wise, J.~H., Abel, T., Turk, M.~J., Norman, M.~L., \& Smith, B.~D. 2012, \mnras, 427, 311, \dodoi{10.1111/j.1365-2966.2012.21809.x}

\bibitem[{Wong {et~al.}(2017)Wong, Hughes, Tokuda, Indebetouw, Bernard, Onishi, Wojciechowski, Bandurski, Kawamura, Roman-Duval, Cao, Chen, Chu, Cui, Fukui, Montier, Muller, Ott, Paradis, Pineda, Rosolowsky, \& Sewiło}]{wong_alma_2017}
Wong, T., Hughes, A., Tokuda, K., {et~al.} 2017, \apj, 850, 139, \dodoi{10.3847/1538-4357/aa9333}

\bibitem[{Wong {et~al.}(2019)Wong, Hughes, Tokuda, Indebetouw, Onishi, Bandurski, Chen, Fukui, Glover, Klessen, Pineda, Roman-Duval, Sewiło, Wojciechowski, \& Zahorecz}]{wong_relations_2019}
---. 2019, \apj, 885, 50, \dodoi{10.3847/1538-4357/ab46ba}

\bibitem[{Wong {et~al.}(2022)Wong, Oudshoorn, Sofovich, Green, Shah, Indebetouw, Meixner, Hacar, Nayak, Tokuda, Bolatto, Chevance, De~Marchi, Fukui, Hirschauer, Jameson, Kalari, Lebouteiller, Looney, Madden, Onishi, Roman-Duval, Rubio, \& Tielens}]{wong_30_2022}
Wong, T., Oudshoorn, L., Sofovich, E., {et~al.} 2022, ApJ, 932, 47, \dodoi{10.3847/1538-4357/ac723a}

\end{thebibliography}
\end{document}